%
%
%
%

\catcode `\@=11 

\def\@version{1.4}
\def\@verdate{22nd Feb 1994}

%
%
%
%


\newif\ifprod@font

\ifx\@typeface\undefined
  \def\@typeface{Comp. Modern}\prod@fontfalse
\else
  \prod@fonttrue 
\fi

\def\newfam{\alloc@8\fam\chardef\sixt@@n} 

\ifprod@font
\font\fiverm=mtr10 at 5pt
\font\fivebf=mtbx10 at 5pt
\font\fiveit=mtti10 at 5pt
\font\fivesl=mtsl10 at 5pt
\font\fivett=mttt10 at 5pt     \hyphenchar\fivett=-1
\font\fivecsc=mtcsc10 at 5pt
\font\fivesf=mtss10 at 5pt
\font\fivei=mtmi10 at 5pt      \skewchar\fivei='177
\font\fivemib=mtmib10 at 5pt   \skewchar\fivemib='177
\font\fivesy=mtsy10 at 5pt     \skewchar\fivesy='60
\font\fivesyb=mtbsy10 at 5pt   \skewchar\fivesyb='60

\font\sixrm=mtr10 at 6pt
\font\sixbf=mtbx10 at 6pt
\font\sixit=mtti10 at 6pt
\font\sixsl=mtsl10 at 6pt
\font\sixtt=mttt10 at 6pt      \hyphenchar\sixtt=-1
\font\sixcsc=mtcsc10 at 6pt
\font\sixsf=mtss10 at 6pt
\font\sixi=mtmi10 at 6pt       \skewchar\sixi='177
\font\sixmib=mtmib10 at 6pt    \skewchar\sixmib='177
\font\sixsy=mtsy10 at 6pt      \skewchar\sixsy='60
\font\sixsyb=mtbsy10 at 6pt    \skewchar\sixsyb='60

\font\sevenrm=mtr10 at 7pt
\font\sevenbf=mtbx10 at 7pt
\font\sevenit=mtti10 at 7pt
\font\sevensl=mtsl10 at 7pt
\font\seventt=mttt10 at 7pt     \hyphenchar\seventt=-1
\font\sevencsc=mtcsc10 at 7pt
\font\sevensf=mtss10 at 7pt
\font\seveni=mtmi10 at 7pt      \skewchar\seveni='177
\font\sevenmib=mtmib10 at 7pt   \skewchar\sevenmib='177
\font\sevensy=mtsy10 at 7pt     \skewchar\sevensy='60
\font\sevensyb=mtbsy10 at 7pt   \skewchar\sevensyb='60

\font\eightrm=mtr10 at 8pt
\font\eightbf=mtbx10 at 8pt
\font\eightit=mtti10 at 8pt
\font\eighti=mtmi10 at 8pt      \skewchar\eighti='177
\font\eightmib=mtmib10 at 8pt   \skewchar\eightmib='177
\font\eightsy=mtsy10 at 8pt     \skewchar\eightsy='60
\font\eightsyb=mtbsy10 at 8pt   \skewchar\eightsyb='60
\font\eightsl=mtsl10 at 8pt
\font\eighttt=mttt10 at 8pt     \hyphenchar\eighttt=-1
\font\eightcsc=mtcsc10 at 8pt
\font\eightsf=mtss10 at 8pt

\font\ninerm=mtr10 at 9pt
\font\ninebf=mtbx10 at 9pt
\font\nineit=mtti10 at 9pt
\font\ninei=mtmi10 at 9pt      \skewchar\ninei='177
\font\ninemib=mtmib10 at 9pt   \skewchar\ninemib='177
\font\ninesy=mtsy10 at 9pt     \skewchar\ninesy='60
\font\ninesyb=mtbsy10 at 9pt   \skewchar\ninesyb='60
\font\ninesl=mtsl10 at 9pt
\font\ninett=mttt10 at 9pt     \hyphenchar\ninett=-1
\font\ninecsc=mtcsc10 at 9pt
\font\ninesf=mtss10 at 9pt

\font\tenrm=mtr10
\font\tenbf=mtbx10
\font\tenit=mtti10
\font\teni=mtmi10		\skewchar\teni='177
\font\tenmib=mtmib10	\skewchar\tenmib='177
\font\tensy=mtsy10		\skewchar\tensy='60
\font\tensyb=mtbsy10	\skewchar\tensyb='60
\font\tenex=cmex10
\font\tensl=mtsl10
\font\tentt=mttt10		\hyphenchar\tentt=-1
\font\tencsc=mtcsc10
\font\tensf=mtss10

\font\elevenrm=mtr10 at 11pt
\font\elevenbf=mtbx10 at 11pt
\font\elevenit=mtti10 at 11pt
\font\eleveni=mtmi10 at 11pt      \skewchar\eleveni='177
\font\elevenmib=mtmib10 at 11pt   \skewchar\elevenmib='177
\font\elevensy=mtsy10 at 11pt     \skewchar\elevensy='60
\font\elevensyb=mtbsy10 at 11pt   \skewchar\elevensyb='60
\font\elevensl=mtsl10 at 11pt
\font\eleventt=mttt10 at 11pt     \hyphenchar\eleventt=-1
\font\elevencsc=mtcsc10 at 11pt
\font\elevensf=mtss10 at 11pt

\font\twelverm=mtr10 at 12pt
\font\twelvebf=mtbx10 at 12pt
\font\twelveit=mtti10 at 12pt
\font\twelvesl=mtsl10 at 12pt
\font\twelvett=mttt10 at 12pt     \hyphenchar\twelvett=-1
\font\twelvecsc=mtcsc10 at 12pt
\font\twelvesf=mtss10 at 12pt
\font\twelvei=mtmi10 at 12pt      \skewchar\twelvei='177
\font\twelvemib=mtmib10 at 12pt   \skewchar\twelvemib='177
\font\twelvesy=mtsy10 at 12pt     \skewchar\twelvesy='60
\font\twelvesyb=mtbsy10 at 12pt   \skewchar\twelvesyb='60

\font\fourteenrm=mtr10 at 14pt
\font\fourteenbf=mtbx10 at 14pt
\font\fourteenit=mtti10 at 14pt
\font\fourteeni=mtmi10 at 14pt      \skewchar\fourteeni='177
\font\fourteenmib=mtmib10 at 14pt   \skewchar\fourteenmib='177
\font\fourteensy=mtsy10 at 14pt     \skewchar\fourteensy='60
\font\fourteensyb=mtbsy10 at 14pt   \skewchar\fourteensyb='60
\font\fourteensl=mtsl10 at 14pt
\font\fourteentt=mttt10 at 14pt     \hyphenchar\fourteentt=-1
\font\fourteencsc=mtcsc10 at 14pt
\font\fourteensf=mtss10 at 14pt

\font\seventeenrm=mtr10 at 17pt
\font\seventeenbf=mtbx10 at 17pt
\font\seventeenit=mtti10 at 17pt
\font\seventeeni=mtmi10 at 17pt      \skewchar\seventeeni='177
\font\seventeenmib=mtmib10 at 17pt   \skewchar\seventeenmib='177
\font\seventeensy=mtsy10 at 17pt     \skewchar\seventeensy='60
\font\seventeensyb=mtbsy10 at 17pt   \skewchar\seventeensyb='60
\font\seventeensl=mtsl10 at 17pt
\font\seventeentt=mttt10 at 17pt     \hyphenchar\seventeentt=-1
\font\seventeencsc=mtcsc10 at 17pt
\font\seventeensf=mtss10 at 17pt


\newfam\xmfam
\newfam\ymfam

\font\fivexm=mtxm10 at 5pt
\font\sixxm=mtxm10 at 6pt
\font\sevenxm=mtxm10 at 7pt
\font\eightxm=mtxm10 at 8pt
\font\ninexm=mtxm10 at 9pt
\font\tenxm=mtxm10
\font\elevenxm=mtxm10 at 11pt
\font\twelvexm=mtxm10 at 12pt
\font\fourteenxm=mtxm10 at 14pt
\font\seventeenxm=mtxm10 at 17pt

\font\fiveym=mtym10 at 5pt
\font\sixym=mtym10 at 6pt
\font\sevenym=mtym10 at 7pt
\font\eightym=mtym10 at 8pt
\font\nineym=mtym10 at 9pt
\font\tenym=mtym10
\font\elevenym=mtym10 at 11pt
\font\twelveym=mtym10 at 12pt
\font\fourteenym=mtym10 at 14pt
\font\seventeenym=mtym10 at 17pt
\else
\font\fiverm=cmr5
\font\fivei=cmmi5             \skewchar\fivei='177
\font\fivemib=cmmib10 at 5pt  \skewchar\fivemib='177
\font\fivesy=cmsy5            \skewchar\fivesy='60
\font\fivesyb=cmbsy10 at 5pt  \skewchar\fivesyb='60
\font\fivebf=cmbx5

\font\sixrm=cmr6
\font\sixi=cmmi6             \skewchar\sixi='177
\font\sixmib=cmmib10 at 6pt  \skewchar\sixmib='177
\font\sixsy=cmsy6            \skewchar\sixsy='60
\font\sixsyb=cmbsy10 at 6pt  \skewchar\sixsyb='60
\font\sixbf=cmbx6

\font\sevenrm=cmr7
\font\seveni=cmmi7             \skewchar\seveni='177
\font\sevenmib=cmmib10 at 7pt  \skewchar\sevenmib='177
\font\sevensy=cmsy7            \skewchar\sevensy='60
\font\sevensyb=cmbsy10 at 7pt  \skewchar\sevensyb='60
\font\sevenbf=cmbx7

\font\eightrm=cmr8
\font\eightbf=cmbx8
\font\eightit=cmti8
\font\eighti=cmmi8			\skewchar\eighti='177
\font\eightmib=cmmib10 at 8pt	\skewchar\eightmib='177
\font\eightsy=cmsy8			\skewchar\eightsy='60
\font\eightsyb=cmbsy10 at 8pt	\skewchar\eightsyb='60
\font\eightsl=cmsl8
\font\eighttt=cmtt8			\hyphenchar\eighttt=-1
\font\eightcsc=cmcsc10 at 8pt
\font\eightsf=cmss8

\font\ninerm=cmr9
\font\ninebf=cmbx9
\font\nineit=cmti9
\font\ninei=cmmi9			\skewchar\ninei='177
\font\ninemib=cmmib10 at 9pt	\skewchar\ninemib='177
\font\ninesy=cmsy9			\skewchar\ninesy='60
\font\ninesyb=cmbsy10 at 9pt	\skewchar\ninesyb='60
\font\ninesl=cmsl9
\font\ninett=cmtt9			\hyphenchar\ninett=-1
\font\ninecsc=cmcsc10 at 9pt
\font\ninesf=cmss9

\font\tenrm=cmr10
\font\tenbf=cmbx10
\font\tenit=cmti10
\font\teni=cmmi10		\skewchar\teni='177
\font\tenmib=cmmib10	\skewchar\tenmib='177
\font\tensy=cmsy10		\skewchar\tensy='60
\font\tensyb=cmbsy10	\skewchar\tensyb='60
\font\tenex=cmex10
\font\tensl=cmsl10
\font\tentt=cmtt10		\hyphenchar\tentt=-1
\font\tencsc=cmcsc10
\font\tensf=cmss10

\font\elevenrm=cmr10 scaled \magstephalf
\font\elevenbf=cmbx10 scaled \magstephalf
\font\elevenit=cmti10 scaled \magstephalf
\font\eleveni=cmmi10 scaled \magstephalf	\skewchar\eleveni='177
\font\elevenmib=cmmib10 scaled \magstephalf	\skewchar\elevenmib='177
\font\elevensy=cmsy10 scaled \magstephalf	\skewchar\elevensy='60
\font\elevensyb=cmbsy10 scaled \magstephalf	\skewchar\elevensyb='60
\font\elevensl=cmsl10 scaled \magstephalf
\font\eleventt=cmtt10 scaled \magstephalf	\hyphenchar\eleventt=-1
\font\elevencsc=cmcsc10 scaled \magstephalf
\font\elevensf=cmss10 scaled \magstephalf

\font\twelverm=cmr10 scaled \magstep1
\font\twelvebf=cmbx10 scaled \magstep1
\font\twelvei=cmmi10 scaled \magstep1      \skewchar\twelvei='177
\font\twelvemib=cmmib10 scaled \magstep1   \skewchar\twelvemib='177
\font\twelvesy=cmsy10 scaled \magstep1     \skewchar\twelvesy='60
\font\twelvesyb=cmbsy10 scaled \magstep1   \skewchar\twelvesyb='60

\font\fourteenrm=cmr10 scaled \magstep2
\font\fourteenbf=cmbx10 scaled \magstep2
\font\fourteenit=cmti10 scaled \magstep2
\font\fourteeni=cmmi10 scaled \magstep2		\skewchar\fourteeni='177
\font\fourteenmib=cmmib10 scaled \magstep2	\skewchar\fourteenmib='177
\font\fourteensy=cmsy10 scaled \magstep2	\skewchar\fourteensy='60
\font\fourteensyb=cmbsy10 scaled \magstep2	\skewchar\fourteensyb='60
\font\fourteensl=cmsl10 scaled \magstep2
\font\fourteentt=cmtt10 scaled \magstep2	\hyphenchar\fourteentt=-1
\font\fourteencsc=cmcsc10 scaled \magstep2
\font\fourteensf=cmss10 scaled \magstep2

\font\seventeenrm=cmr10 scaled \magstep3
\font\seventeenbf=cmbx10 scaled \magstep3
\font\seventeenit=cmti10 scaled \magstep3
\font\seventeeni=cmmi10 scaled \magstep3	\skewchar\seventeeni='177
\font\seventeenmib=cmmib10 scaled \magstep3	\skewchar\seventeenmib='177
\font\seventeensy=cmsy10 scaled \magstep3	\skewchar\seventeensy='60
\font\seventeensyb=cmbsy10 scaled \magstep3	\skewchar\seventeensyb='60
\font\seventeensl=cmsl10 scaled \magstep3
\font\seventeentt=cmtt10 scaled \magstep3	\hyphenchar\seventeentt=-1
\font\seventeencsc=cmcsc10 scaled \magstep3
\font\seventeensf=cmss10 scaled \magstep3
\fi

\def\hexnumber#1{\ifcase#1 0\or1\or2\or3\or4\or5\or6\or7\or8\or9\or
  A\or B\or C\or D\or E\or F\fi}

\ifprod@font
  \edef\@xm{\hexnumber\xmfam}
  \edef\@ym{\hexnumber\ymfam}
\fi

\def\makestrut{%
  \setbox\strutbox=\hbox{%
    \vrule height.7\baselineskip depth.3\baselineskip width \z@}%
}

\def\baselinestretch{1}
\newskip\tmp@bls

\def\b@ls#1{
  \tmp@bls=#1\relax
  \baselineskip=#1\relax\makestrut
  \normalbaselineskip=\baselinestretch\tmp@bls
  \normalbaselines
}

\def\nostb@ls#1{
  \normalbaselineskip=#1\relax
  \normalbaselines
  \makestrut
}

%

\newfam\mibfam 
\newfam\sybfam 
\newfam\scfam  
\newfam\sffam  

\def\mit{\fam\@ne}

\def\cal{\fam\tw@}

\def\em{\ifdim\fontdimen1\font>\z@ \rm\else\it\fi}

\textfont3=\tenex
\scriptfont3=\tenex
\scriptscriptfont3=\tenex

\setbox0=\hbox{\tenex B} \p@renwd=\wd0 

\def\eightpoint{
  \def\rm{\fam0\eightrm}%
  \textfont0=\eightrm \scriptfont0=\sixrm \scriptscriptfont0=\fiverm%
  \textfont1=\eighti  \scriptfont1=\sixi  \scriptscriptfont1=\fivei%
  \textfont2=\eightsy \scriptfont2=\sixsy \scriptscriptfont2=\fivesy%
  \textfont\itfam=\eightit\def\it{\fam\itfam\eightit}%
  \ifprod@font
    \scriptfont\itfam=\sixit
      \scriptscriptfont\itfam=\fiveit
  \else
    \scriptfont\itfam=\eightit
      \scriptscriptfont\itfam=\eightit
  \fi
  \textfont\bffam=\eightbf%
    \scriptfont\bffam=\sixbf%
      \scriptscriptfont\bffam=\fivebf%
  \def\bf{\fam\bffam\eightbf}%
  \textfont\slfam=\eightsl\def\sl{\fam\slfam\eightsl}%
  \ifprod@font
    \scriptfont\slfam=\sixsl
      \scriptscriptfont\slfam=\fivesl
  \else
    \scriptfont\slfam=\eightsl
      \scriptscriptfont\slfam=\eightsl
  \fi
  \textfont\ttfam=\eighttt\def\tt{\fam\ttfam\eighttt}%
  \ifprod@font
    \scriptfont\ttfam=\sixtt
      \scriptscriptfont\ttfam=\fivett
  \else
    \scriptfont\ttfam=\eighttt
      \scriptscriptfont\ttfam=\eighttt
  \fi
  \textfont\scfam=\eightcsc\def\sc{\fam\scfam\eightcsc}%
  \ifprod@font
    \scriptfont\scfam=\sixcsc
      \scriptscriptfont\scfam=\fivecsc
  \else
    \scriptfont\scfam=\eightcsc
      \scriptscriptfont\scfam=\eightcsc
  \fi
  \textfont\sffam=\eightsf\def\sf{\fam\sffam\eightsf}%
  \ifprod@font
    \scriptfont\sffam=\sixsf
      \scriptscriptfont\sffam=\fivesf
  \else
    \scriptfont\sffam=\eightsf
      \scriptscriptfont\sffam=\eightsf
  \fi
  \textfont\mibfam=\eightmib
    \scriptfont\mibfam=\sixmib
      \scriptscriptfont\mibfam=\fivemib
  \textfont\sybfam=\eightsyb
    \scriptfont\sybfam=\sixsyb
      \scriptscriptfont\sybfam=\fivesyb
  \ifprod@font
    \textfont\xmfam=\eightxm
      \scriptfont\xmfam=\sixxm
        \scriptscriptfont\xmfam=\fivexm
    \textfont\ymfam=\eightym
      \scriptfont\ymfam=\sixym
        \scriptscriptfont\ymfam=\fiveym
  \fi
  \def\oldstyle{\fam\@ne\eighti}%
  \def\boldstyle{\fam\mibfam\eightmib}%
  \b@ls{10pt}\rm%
}

\def\ninepoint{
  \def\rm{\fam0\ninerm}%
  \textfont0=\ninerm \scriptfont0=\sixrm \scriptscriptfont0=\fiverm%
  \textfont1=\ninei  \scriptfont1=\sixi  \scriptscriptfont1=\fivei%
  \textfont2=\ninesy \scriptfont2=\sixsy \scriptscriptfont2=\fivesy%
  \textfont\itfam=\nineit\def\it{\fam\itfam\nineit}%
  \ifprod@font
    \scriptfont\itfam=\sixit
      \scriptscriptfont\itfam=\fiveit
  \else
    \scriptfont\itfam=\nineit
      \scriptscriptfont\itfam=\nineit
  \fi
  \textfont\bffam=\ninebf%
    \scriptfont\bffam=\sixbf%
      \scriptscriptfont\bffam=\fivebf%
  \def\bf{\fam\bffam\ninebf}%
  \textfont\slfam=\ninesl\def\sl{\fam\slfam\ninesl}%
  \ifprod@font
    \scriptfont\slfam=\sixsl
      \scriptscriptfont\slfam=\fivesl
  \else
    \scriptfont\slfam=\ninesl
      \scriptscriptfont\slfam=\ninesl
  \fi
  \textfont\ttfam=\ninett\def\tt{\fam\ttfam\ninett}%
  \ifprod@font
    \scriptfont\ttfam=\sixtt
      \scriptscriptfont\ttfam=\fivett
  \else
    \scriptfont\ttfam=\ninett
      \scriptscriptfont\ttfam=\ninett
  \fi
  \textfont\scfam=\ninecsc\def\sc{\fam\scfam\ninecsc}%
  \ifprod@font
    \scriptfont\scfam=\sixcsc
      \scriptscriptfont\scfam=\fivecsc
  \else
    \scriptfont\scfam=\ninecsc
      \scriptscriptfont\scfam=\ninecsc
  \fi
  \textfont\sffam=\ninesf\def\sf{\fam\sffam\ninesf}%
  \ifprod@font
    \scriptfont\sffam=\sixsf
      \scriptscriptfont\sffam=\fivesf
  \else
    \scriptfont\sffam=\ninesf
      \scriptscriptfont\sffam=\ninesf
  \fi
  \textfont\mibfam=\ninemib
    \scriptfont\mibfam=\sixmib
      \scriptscriptfont\mibfam=\fivemib
  \textfont\sybfam=\ninesyb
    \scriptfont\sybfam=\sixsyb
      \scriptscriptfont\sybfam=\fivesyb
  \ifprod@font
    \textfont\xmfam=\ninexm
      \scriptfont\xmfam=\sixxm
        \scriptscriptfont\xmfam=\fivexm
    \textfont\ymfam=\nineym
      \scriptfont\ymfam=\sixym
        \scriptscriptfont\ymfam=\fiveym
  \fi
  \def\oldstyle{\fam\@ne\ninei}%
  \def\boldstyle{\fam\mibfam\ninemib}%
  \b@ls{\TextLeading plus \Feathering}\rm%
}

\def\tenpoint{
  \def\rm{\fam0\tenrm}%
  \textfont0=\tenrm \scriptfont0=\sevenrm \scriptscriptfont0=\fiverm%
  \textfont1=\teni  \scriptfont1=\seveni  \scriptscriptfont1=\fivei%
  \textfont2=\tensy \scriptfont2=\sevensy \scriptscriptfont2=\fivesy%
  \textfont\itfam=\tenit\def\it{\fam\itfam\tenit}%
  \ifprod@font
    \scriptfont\itfam=\sevenit
      \scriptscriptfont\itfam=\fiveit
  \else
    \scriptfont\itfam=\tenit
      \scriptscriptfont\itfam=\tenit
  \fi
  \textfont\bffam=\tenbf%
    \scriptfont\bffam=\sevenbf%
      \scriptscriptfont\bffam=\fivebf%
  \def\bf{\fam\bffam\tenbf}%
  \textfont\slfam=\tensl\def\sl{\fam\slfam\tensl}%
  \ifprod@font
    \scriptfont\slfam=\sevensl
      \scriptscriptfont\slfam=\fivesl
  \else
    \scriptfont\slfam=\tensl
      \scriptscriptfont\slfam=\tensl
  \fi
  \textfont\ttfam=\tentt\def\tt{\fam\ttfam\tentt}%
  \ifprod@font
    \scriptfont\ttfam=\seventt
      \scriptscriptfont\ttfam=\fivett
  \else
    \scriptfont\ttfam=\tentt
      \scriptscriptfont\ttfam=\tentt
  \fi
  \textfont\scfam=\tencsc\def\sc{\fam\scfam\tencsc}%
  \ifprod@font
    \scriptfont\scfam=\sevencsc
      \scriptscriptfont\scfam=\fivecsc
  \else
    \scriptfont\scfam=\tencsc
      \scriptscriptfont\scfam=\tencsc
  \fi
  \textfont\sffam=\tensf\def\sf{\fam\sffam\tensf}%
  \ifprod@font
    \scriptfont\sffam=\sevensf
      \scriptscriptfont\sffam=\fivesf
  \else
    \scriptfont\sffam=\tensf
      \scriptscriptfont\sffam=\tensf
  \fi
  \textfont\mibfam=\tenmib
    \scriptfont\mibfam=\sevenmib
      \scriptscriptfont\mibfam=\fivemib
  \textfont\sybfam=\tensyb
    \scriptfont\sybfam=\sevensyb
      \scriptscriptfont\sybfam=\fivesyb
  \ifprod@font
    \textfont\xmfam=\tenxm
      \scriptfont\xmfam=\sevenxm
        \scriptscriptfont\xmfam=\fivexm
    \textfont\ymfam=\tenym
      \scriptfont\ymfam=\sevenym
        \scriptscriptfont\ymfam=\fiveym
  \fi
  \def\oldstyle{\fam\@ne\teni}%
  \def\boldstyle{\fam\mibfam\tenmib}%
  \b@ls{11pt}\rm%
}

\def\elevenpoint{
  \def\rm{\fam0\elevenrm}%
  \textfont0=\elevenrm \scriptfont0=\eightrm \scriptscriptfont0=\sixrm%
  \textfont1=\eleveni  \scriptfont1=\eighti  \scriptscriptfont1=\sixi%
  \textfont2=\elevensy \scriptfont2=\eightsy \scriptscriptfont2=\sixsy%
  \textfont\itfam=\elevenit\def\it{\fam\itfam\elevenit}%
  \ifprod@font
    \scriptfont\itfam=\eightit
      \scriptscriptfont\itfam=\sixit
  \else
    \scriptfont\itfam=\elevenit
      \scriptscriptfont\itfam=\elevenit
  \fi
  \textfont\bffam=\elevenbf%
    \scriptfont\bffam=\eightbf%
      \scriptscriptfont\bffam=\sixbf%
  \def\bf{\fam\bffam\elevenbf}%
  \textfont\slfam=\elevensl\def\sl{\fam\slfam\elevensl}%
  \ifprod@font
    \scriptfont\slfam=\eightsl
      \scriptscriptfont\slfam=\sixsl
  \else
    \scriptfont\slfam=\elevensl
      \scriptscriptfont\slfam=\elevensl
  \fi
  \textfont\ttfam=\eleventt\def\tt{\fam\ttfam\eleventt}%
  \ifprod@font
    \scriptfont\ttfam=\eighttt
      \scriptscriptfont\ttfam=\sixtt
  \else
    \scriptfont\ttfam=\eleventt
      \scriptscriptfont\ttfam=\eleventt
  \fi
  \textfont\scfam=\elevencsc\def\sc{\fam\scfam\elevencsc}%
  \ifprod@font
    \scriptfont\scfam=\eightcsc
      \scriptscriptfont\scfam=\sixcsc
  \else
    \scriptfont\scfam=\elevencsc
      \scriptscriptfont\scfam=\elevencsc
  \fi
  \textfont\sffam=\elevensf\def\sf{\fam\sffam\elevensf}%
  \ifprod@font
    \scriptfont\sffam=\eightsf
      \scriptscriptfont\sffam=\sixsf
  \else
    \scriptfont\sffam=\elevensf
      \scriptscriptfont\sffam=\elevensf
  \fi
  \textfont\mibfam=\elevenmib
    \scriptfont\mibfam=\eightmib
      \scriptscriptfont\mibfam=\sixmib
  \textfont\sybfam=\elevensyb
    \scriptfont\sybfam=\eightsyb
      \scriptscriptfont\sybfam=\sixsyb
  \ifprod@font
    \textfont\xmfam=\elevenxm
      \scriptfont\xmfam=\eightxm
       \scriptscriptfont\xmfam=\sixxm
    \textfont\ymfam=\elevenym
      \scriptfont\ymfam=\eightym
        \scriptscriptfont\ymfam=\sixym
   \fi
  \def\oldstyle{\fam\@ne\eleveni}%
  \def\boldstyle{\fam\mibfam\elevenmib}%
  \b@ls{13pt}\rm%
}

\def\fourteenpoint{
  \def\rm{\fam0\fourteenrm}%
  \textfont0\fourteenrm  \scriptfont0\tenrm  \scriptscriptfont0\sevenrm%
  \textfont1\fourteeni   \scriptfont1\teni   \scriptscriptfont1\seveni%
  \textfont2\fourteensy  \scriptfont2\tensy  \scriptscriptfont2\sevensy%
  \textfont\itfam=\fourteenit\def\it{\fam\itfam\fourteenit}%
  \ifprod@font
    \scriptfont\itfam=\tenit
      \scriptscriptfont\itfam=\sevenit
  \else
    \scriptfont\itfam=\fourteenit
      \scriptscriptfont\itfam=\fourteenit
  \fi
  \textfont\bffam=\fourteenbf%
    \scriptfont\bffam=\tenbf%
      \scriptscriptfont\bffam=\sevenbf%
  \def\bf{\fam\bffam\fourteenbf}%
  \textfont\slfam=\fourteensl\def\sl{\fam\slfam\fourteensl}%
  \ifprod@font
    \scriptfont\slfam=\tensl
      \scriptscriptfont\slfam=\sevensl
  \else
    \scriptfont\slfam=\fourteensl
      \scriptscriptfont\slfam=\fourteensl
  \fi
  \textfont\ttfam=\fourteentt\def\tt{\fam\ttfam\fourteentt}%
  \ifprod@font
    \scriptfont\ttfam=\tentt
      \scriptscriptfont\ttfam=\seventt
  \else
    \scriptfont\ttfam=\fourteentt
      \scriptscriptfont\ttfam=\fourteentt
  \fi
  \textfont\scfam=\fourteencsc\def\sc{\fam\scfam\fourteencsc}%
  \ifprod@font
    \scriptfont\scfam=\tencsc
      \scriptscriptfont\scfam=\sevencsc
  \else
    \scriptfont\scfam=\fourteencsc
      \scriptscriptfont\scfam=\fourteencsc
  \fi
  \textfont\sffam=\fourteensf\def\sf{\fam\sffam\fourteensf}%
  \ifprod@font
    \scriptfont\sffam=\tensf
      \scriptscriptfont\sffam=\sevensf
  \else
    \scriptfont\sffam=\fourteensf
      \scriptscriptfont\sffam=\fourteensf
  \fi
  \textfont\mibfam=\fourteenmib
    \scriptfont\mibfam=\tenmib
      \scriptscriptfont\mibfam=\sevenmib
  \textfont\sybfam=\fourteensyb
    \scriptfont\sybfam=\tensyb
      \scriptscriptfont\sybfam=\sevensyb
  \ifprod@font
    \textfont\xmfam=\fourteenxm
      \scriptfont\xmfam=\tenxm
        \scriptscriptfont\xmfam=\sevenxm
   \textfont\ymfam=\fourteenym
      \scriptfont\ymfam=\tenym
        \scriptscriptfont\ymfam=\sevenym
  \fi
  \def\oldstyle{\fam\@ne\fourteeni}%
  \def\boldstyle{\fam\mibfam\fourteenmib}%
  \b@ls{17pt}\rm%
}

\def\seventeenpoint{
  \def\rm{\fam0\seventeenrm}%
  \textfont0\seventeenrm  \scriptfont0\twelverm  \scriptscriptfont0\tenrm%
  \textfont1\seventeeni   \scriptfont1\twelvei   \scriptscriptfont1\teni%
  \textfont2\seventeensy  \scriptfont2\twelvesy  \scriptscriptfont2\tensy%
  \textfont\itfam=\seventeenit\def\it{\fam\itfam\seventeenit}%
  \ifprod@font
    \scriptfont\itfam=\twelveit
      \scriptscriptfont\itfam=\tenit
  \else
    \scriptfont\itfam=\seventeenit
      \scriptscriptfont\itfam=\seventeenit
  \fi
  \textfont\bffam=\seventeenbf%
    \scriptfont\bffam=\twelvebf%
      \scriptscriptfont\bffam=\tenbf%
  \def\bf{\fam\bffam\seventeenbf}%
  \textfont\slfam=\seventeensl\def\sl{\fam\slfam\seventeensl}%
  \ifprod@font
    \scriptfont\slfam=\twelvesl
      \scriptscriptfont\slfam=\tensl
  \else
    \scriptfont\slfam=\seventeensl
      \scriptscriptfont\slfam=\seventeensl
  \fi
  \textfont\ttfam=\seventeentt\def\tt{\fam\ttfam\seventeentt}%
  \ifprod@font
    \scriptfont\ttfam=\twelvett
      \scriptscriptfont\ttfam=\tentt
  \else
    \scriptfont\ttfam=\seventeentt
      \scriptscriptfont\ttfam=\seventeentt
  \fi
  \textfont\scfam=\seventeencsc\def\sc{\fam\scfam\seventeencsc}%
  \ifprod@font
    \scriptfont\scfam=\twelvecsc
      \scriptscriptfont\scfam=\tencsc
  \else
    \scriptfont\scfam=\seventeencsc
      \scriptscriptfont\scfam=\seventeencsc
  \fi
  \textfont\sffam=\seventeensf\def\sf{\fam\sffam\seventeensf}%
  \ifprod@font
    \scriptfont\sffam=\twelvesf
      \scriptscriptfont\sffam=\tensf
  \else
    \scriptfont\sffam=\seventeensf
      \scriptscriptfont\sffam=\seventeensf
  \fi
  \textfont\mibfam=\seventeenmib
    \scriptfont\mibfam=\twelvemib
      \scriptscriptfont\mibfam=\tenmib
  \textfont\sybfam=\seventeensyb
    \scriptfont\sybfam=\twelvesyb
      \scriptscriptfont\sybfam=\tensyb
  \ifprod@font
    \textfont\xmfam=\seventeenxm
      \scriptfont\xmfam=\twelvexm
        \scriptscriptfont\xmfam=\tenxm
    \textfont\ymfam=\seventeenym
      \scriptfont\ymfam=\twelveym
        \scriptscriptfont\ymfam=\tenym
  \fi
  \def\oldstyle{\fam\@ne\seventeeni}%
  \def\boldstyle{\fam\mibfam\seventeenmib}%
  \b@ls{20pt}\rm%
}

\lineskip=1pt      \normallineskip=\lineskip
\lineskiplimit=\z@ \normallineskiplimit=\lineskiplimit



\def\la{\mathrel{\mathchoice {\vcenter{\offinterlineskip\halign{\hfil
$\displaystyle##$\hfil\cr<\cr\sim\cr}}}
{\vcenter{\offinterlineskip\halign{\hfil$\textstyle##$\hfil\cr
<\cr\sim\cr}}}
{\vcenter{\offinterlineskip\halign{\hfil$\scriptstyle##$\hfil\cr
<\cr\sim\cr}}}
{\vcenter{\offinterlineskip\halign{\hfil$\scriptscriptstyle##$\hfil\cr
<\cr\sim\cr}}}}}

\def\ga{\mathrel{\mathchoice {\vcenter{\offinterlineskip\halign{\hfil
$\displaystyle##$\hfil\cr>\cr\sim\cr}}}
{\vcenter{\offinterlineskip\halign{\hfil$\textstyle##$\hfil\cr
>\cr\sim\cr}}}
{\vcenter{\offinterlineskip\halign{\hfil$\scriptstyle##$\hfil\cr
>\cr\sim\cr}}}
{\vcenter{\offinterlineskip\halign{\hfil$\scriptscriptstyle##$\hfil\cr
>\cr\sim\cr}}}}}

\def\getsto{\mathrel{\mathchoice {\vcenter{\offinterlineskip
\halign{\hfil
$\displaystyle##$\hfil\cr\gets\cr\to\cr}}}
{\vcenter{\offinterlineskip\halign{\hfil$\textstyle##$\hfil\cr\gets
\cr\to\cr}}}
{\vcenter{\offinterlineskip\halign{\hfil$\scriptstyle##$\hfil\cr\gets
\cr\to\cr}}}
{\vcenter{\offinterlineskip\halign{\hfil$\scriptscriptstyle##$\hfil\cr
\gets\cr\to\cr}}}}}

\def\lid{\mathrel{\mathchoice {\vcenter{\offinterlineskip\halign{\hfil
$\displaystyle##$\hfil\cr<\cr\noalign{\vskip1.2pt}=\cr}}}
{\vcenter{\offinterlineskip\halign{\hfil$\textstyle##$\hfil\cr<\cr
\noalign{\vskip1.2pt}=\cr}}}
{\vcenter{\offinterlineskip\halign{\hfil$\scriptstyle##$\hfil\cr<\cr
\noalign{\vskip1pt}=\cr}}}
{\vcenter{\offinterlineskip\halign{\hfil$\scriptscriptstyle##$\hfil\cr
<\cr
\noalign{\vskip0.9pt}=\cr}}}}}

\def\gid{\mathrel{\mathchoice {\vcenter{\offinterlineskip\halign{\hfil
$\displaystyle##$\hfil\cr>\cr\noalign{\vskip1.2pt}=\cr}}}
{\vcenter{\offinterlineskip\halign{\hfil$\textstyle##$\hfil\cr>\cr
\noalign{\vskip1.2pt}=\cr}}}
{\vcenter{\offinterlineskip\halign{\hfil$\scriptstyle##$\hfil\cr>\cr
\noalign{\vskip1pt}=\cr}}}
{\vcenter{\offinterlineskip\halign{\hfil$\scriptscriptstyle##$\hfil\cr
>\cr
\noalign{\vskip0.9pt}=\cr}}}}}

\def\grole{\mathrel{\mathchoice {\vcenter{\offinterlineskip\halign{\hfil
$\displaystyle##$\hfil\cr>\cr\noalign{\vskip-1.5pt}<\cr}}}
{\vcenter{\offinterlineskip\halign{\hfil$\textstyle##$\hfil\cr
>\cr\noalign{\vskip-1.5pt}<\cr}}}
{\vcenter{\offinterlineskip\halign{\hfil$\scriptstyle##$\hfil\cr
>\cr\noalign{\vskip-1pt}<\cr}}}
{\vcenter{\offinterlineskip\halign{\hfil$\scriptscriptstyle##$\hfil\cr
>\cr\noalign{\vskip-0.5pt}<\cr}}}}}

\def\leogr{\mathrel{\mathchoice {\vcenter{\offinterlineskip\halign{\hfil
$\displaystyle##$\hfil\cr<\cr\noalign{\vskip-1.5pt}>\cr}}}
{\vcenter{\offinterlineskip\halign{\hfil$\textstyle##$\hfil\cr
<\cr\noalign{\vskip-1.5pt}>\cr}}}
{\vcenter{\offinterlineskip\halign{\hfil$\scriptstyle##$\hfil\cr
<\cr\noalign{\vskip-1pt}>\cr}}}
{\vcenter{\offinterlineskip\halign{\hfil$\scriptscriptstyle##$\hfil\cr
<\cr\noalign{\vskip-0.5pt}>\cr}}}}}

\def\loa{\mathrel{\mathchoice {\vcenter{\offinterlineskip\halign{\hfil
$\displaystyle##$\hfil\cr<\cr\approx\cr}}}
{\vcenter{\offinterlineskip\halign{\hfil$\textstyle##$\hfil\cr
<\cr\approx\cr}}}
{\vcenter{\offinterlineskip\halign{\hfil$\scriptstyle##$\hfil\cr
<\cr\approx\cr}}}
{\vcenter{\offinterlineskip\halign{\hfil$\scriptscriptstyle##$\hfil\cr
<\cr\approx\cr}}}}}

\def\goa{\mathrel{\mathchoice {\vcenter{\offinterlineskip\halign{\hfil
$\displaystyle##$\hfil\cr>\cr\approx\cr}}}
{\vcenter{\offinterlineskip\halign{\hfil$\textstyle##$\hfil\cr
>\cr\approx\cr}}}
{\vcenter{\offinterlineskip\halign{\hfil$\scriptstyle##$\hfil\cr
>\cr\approx\cr}}}
{\vcenter{\offinterlineskip\halign{\hfil$\scriptscriptstyle##$\hfil\cr
>\cr\approx\cr}}}}}

\def\diameter{{\ifmmode\mathchoice
{\ooalign{\hfil\hbox{$\displaystyle/$}\hfil\crcr
{\hbox{$\displaystyle\mathchar"20D$}}}}
{\ooalign{\hfil\hbox{$\textstyle/$}\hfil\crcr
{\hbox{$\textstyle\mathchar"20D$}}}}
{\ooalign{\hfil\hbox{$\scriptstyle/$}\hfil\crcr
{\hbox{$\scriptstyle\mathchar"20D$}}}}
{\ooalign{\hfil\hbox{$\scriptscriptstyle/$}\hfil\crcr
{\hbox{$\scriptscriptstyle\mathchar"20D$}}}}
\else{\ooalign{\hfil/\hfil\crcr\mathhexbox20D}}%
\fi}}

\def\sq{\ifmmode\squareforqed\else{\unskip\nobreak\hfil
\penalty50\hskip1em\null\nobreak\hfil\squareforqed
\parfillskip=0pt\finalhyphendemerits=0\endgraf}\fi}
\def\squareforqed{\hbox{\rlap{$\sqcap$}$\sqcup$}}


\def\bbbc{{\mathchoice {\setbox0=\hbox{$\displaystyle\rm C$}\hbox{\hbox
to0pt{\kern0.4\wd0\vrule height0.9\ht0\hss}\box0}}
{\setbox0=\hbox{$\textstyle\rm C$}\hbox{\hbox
to0pt{\kern0.4\wd0\vrule height0.9\ht0\hss}\box0}}
{\setbox0=\hbox{$\scriptstyle\rm C$}\hbox{\hbox
to0pt{\kern0.4\wd0\vrule height0.9\ht0\hss}\box0}}
{\setbox0=\hbox{$\scriptscriptstyle\rm C$}\hbox{\hbox
to0pt{\kern0.4\wd0\vrule height0.9\ht0\hss}\box0}}}}
\def\bbbq{{\mathchoice {\setbox0=\hbox{$\displaystyle\rm
Q$}\hbox{\raise
0.15\ht0\hbox to0pt{\kern0.4\wd0\vrule height0.8\ht0\hss}\box0}}
{\setbox0=\hbox{$\textstyle\rm Q$}\hbox{\raise
0.15\ht0\hbox to0pt{\kern0.4\wd0\vrule height0.8\ht0\hss}\box0}}
{\setbox0=\hbox{$\scriptstyle\rm Q$}\hbox{\raise
0.15\ht0\hbox to0pt{\kern0.4\wd0\vrule height0.7\ht0\hss}\box0}}
{\setbox0=\hbox{$\scriptscriptstyle\rm Q$}\hbox{\raise
0.15\ht0\hbox to0pt{\kern0.4\wd0\vrule height0.7\ht0\hss}\box0}}}}
\def\bbbt{{\mathchoice {\setbox0=\hbox{$\displaystyle\rm
T$}\hbox{\hbox to0pt{\kern0.3\wd0\vrule height0.9\ht0\hss}\box0}}
{\setbox0=\hbox{$\textstyle\rm T$}\hbox{\hbox
to0pt{\kern0.3\wd0\vrule height0.9\ht0\hss}\box0}}
{\setbox0=\hbox{$\scriptstyle\rm T$}\hbox{\hbox
to0pt{\kern0.3\wd0\vrule height0.9\ht0\hss}\box0}}
{\setbox0=\hbox{$\scriptscriptstyle\rm T$}\hbox{\hbox
to0pt{\kern0.3\wd0\vrule height0.9\ht0\hss}\box0}}}}
\def\bbbs{{\mathchoice
{\setbox0=\hbox{$\displaystyle     \rm S$}\hbox{\raise0.5\ht0\hbox
to0pt{\kern0.35\wd0\vrule height0.45\ht0\hss}\hbox
to0pt{\kern0.55\wd0\vrule height0.5\ht0\hss}\box0}}
{\setbox0=\hbox{$\textstyle        \rm S$}\hbox{\raise0.5\ht0\hbox
to0pt{\kern0.35\wd0\vrule height0.45\ht0\hss}\hbox
to0pt{\kern0.55\wd0\vrule height0.5\ht0\hss}\box0}}
{\setbox0=\hbox{$\scriptstyle      \rm S$}\hbox{\raise0.5\ht0\hbox
to0pt{\kern0.35\wd0\vrule height0.45\ht0\hss}\raise0.05\ht0\hbox
to0pt{\kern0.5\wd0\vrule height0.45\ht0\hss}\box0}}
{\setbox0=\hbox{$\scriptscriptstyle\rm S$}\hbox{\raise0.5\ht0\hbox
to0pt{\kern0.4\wd0\vrule height0.45\ht0\hss}\raise0.05\ht0\hbox
to0pt{\kern0.55\wd0\vrule height0.45\ht0\hss}\box0}}}}
\def\bbbz{{\mathchoice {\hbox{$\sf\textstyle Z\kern-0.4em Z$}}
{\hbox{$\sf\textstyle Z\kern-0.4em Z$}}
{\hbox{$\sf\scriptstyle Z\kern-0.3em Z$}}
{\hbox{$\sf\scriptscriptstyle Z\kern-0.2em Z$}}}}


\ifprod@font
  \mathchardef\la="3\@xm2E
  \mathchardef\getsto="3\@xm1C
  \mathchardef\lid="3\@xm35
  \mathchardef\grole="3\@xm3F
  \mathchardef\loa="3\@xm2F
  \mathchardef\ga="3\@xm26
  \mathchardef\gid="3\@xm3D
  \mathchardef\leogr="3\@xm37
  \mathchardef\goa="3\@xm27
  \mathchardef\sq="0\@xm03
%
%
\def\diameter{{%
  \ifmmode
    \mathchoice
    {\ooalign{\hfil\hbox{$\displaystyle/$}\hfil\crcr
    {\lower.2ex\hbox{$\displaystyle\mathchar"20D$}}}}%
    {\ooalign{\hfil\hbox{$\textstyle/$}\hfil\crcr
    {\lower.2ex\hbox{$\textstyle\mathchar"20D$}}}}%
    {\ooalign{\hfil\hbox{$\scriptstyle/$}\hfil\crcr
    {\lower.1ex\hbox{$\scriptstyle\mathchar"20D$}}}}%
    {\ooalign{\hfil\hbox{$\scriptscriptstyle/$}\hfil\crcr
    {\lower.1ex\hbox{$\scriptscriptstyle\mathchar"20D$}}}}%
  \else
    {\ooalign{\hfil/\hfil\crcr\lower.2ex\hbox{\mathhexbox20D}}}%
  \fi
}}
%
%

\def\bbbc{{\Bbb{C}}}
\def\bbbq{{\Bbb{Q}}}
\def\bbbt{{\Bbb{T}}}
\def\bbbs{{\Bbb{S}}}
\def\bbbz{{\Bbb{Z}}}
\fi


\ifprod@font
\mathchardef\boxdot="2\@xm00
\mathchardef\boxplus="2\@xm01
\mathchardef\boxtimes="2\@xm02
\mathchardef\square="0\@xm03
\mathchardef\blacksquare="0\@xm04
\mathchardef\centerdot="2\@xm05
\mathchardef\lozenge="0\@xm06
\mathchardef\blacklozenge="0\@xm07
\mathchardef\circlearrowright="3\@xm08
\mathchardef\circlearrowleft="3\@xm09
\mathchardef\rightleftharpoons="3\@xm0A
\mathchardef\leftrightharpoons="3\@xm0B
\mathchardef\boxminus="2\@xm0C
\mathchardef\Vdash="3\@xm0D
\mathchardef\Vvdash="3\@xm0E
\mathchardef\vDash="3\@xm0F
\mathchardef\twoheadrightarrow="3\@xm10
\mathchardef\twoheadleftarrow="3\@xm11
\mathchardef\leftleftarrows="3\@xm12
\mathchardef\rightrightarrows="3\@xm13
\mathchardef\upuparrows="3\@xm14
\mathchardef\downdownarrows="3\@xm15
\mathchardef\upharpoonright="3\@xm16

\mathchardef\downharpoonright="3\@xm17
\mathchardef\upharpoonleft="3\@xm18
\mathchardef\downharpoonleft="3\@xm19
\mathchardef\rightarrowtail="3\@xm1A
\mathchardef\leftarrowtail="3\@xm1B
\mathchardef\leftrightarrows="3\@xm1C
\mathchardef\rightleftarrows="3\@xm1D
\mathchardef\Lsh="3\@xm1E
\mathchardef\Rsh="3\@xm1F
\mathchardef\rightsquigarrow="3\@xm20
\mathchardef\leftrightsquigarrow="3\@xm21
\mathchardef\looparrowleft="3\@xm22
\mathchardef\looparrowright="3\@xm23
\mathchardef\circeq="3\@xm24
\mathchardef\succsim="3\@xm25
\mathchardef\gtrsim="3\@xm26
\mathchardef\gtrapprox="3\@xm27
\mathchardef\multimap="3\@xm28
\mathchardef\therefore="3\@xm29
\mathchardef\because="3\@xm2A
\mathchardef\doteqdot="3\@xm2B

\mathchardef\triangleq="3\@xm2C
\mathchardef\precsim="3\@xm2D
\mathchardef\lesssim="3\@xm2E
\mathchardef\lessapprox="3\@xm2F
\mathchardef\eqslantless="3\@xm30
\mathchardef\eqslantgtr="3\@xm31
\mathchardef\curlyeqprec="3\@xm32
\mathchardef\curlyeqsucc="3\@xm33
\mathchardef\preccurlyeq="3\@xm34
\mathchardef\leqq="3\@xm35
\mathchardef\leqslant="3\@xm36
\mathchardef\lessgtr="3\@xm37
\mathchardef\backprime="0\@xm38
\mathchardef\risingdotseq="3\@xm3A
\mathchardef\fallingdotseq="3\@xm3B
\mathchardef\succcurlyeq="3\@xm3C
\mathchardef\geqq="3\@xm3D
\mathchardef\geqslant="3\@xm3E
\mathchardef\gtrless="3\@xm3F
\mathchardef\sqsubset="3\@xm40
\mathchardef\sqsupset="3\@xm41
\mathchardef\vartriangleright="3\@xm42
\mathchardef\vartriangleleft="3\@xm43
\mathchardef\trianglerighteq="3\@xm44
\mathchardef\trianglelefteq="3\@xm45
\mathchardef\bigstar="0\@xm46
\mathchardef\between="3\@xm47
\mathchardef\blacktriangledown="0\@xm48
\mathchardef\blacktriangleright="3\@xm49
\mathchardef\blacktriangleleft="3\@xm4A
\mathchardef\vartriangle="0\@xm4D
\mathchardef\blacktriangle="0\@xm4E
\mathchardef\triangledown="0\@xm4F
\mathchardef\eqcirc="3\@xm50
\mathchardef\lesseqgtr="3\@xm51
\mathchardef\gtreqless="3\@xm52
\mathchardef\lesseqqgtr="3\@xm53
\mathchardef\gtreqqless="3\@xm54
\mathchardef\Rrightarrow="3\@xm56
\mathchardef\Lleftarrow="3\@xm57
\mathchardef\veebar="2\@xm59
\mathchardef\barwedge="2\@xm5A
\mathchardef\doublebarwedge="2\@xm5B
\mathchardef\angle="0\@xm5C
\mathchardef\measuredangle="0\@xm5D
\mathchardef\sphericalangle="0\@xm5E
\mathchardef\varpropto="3\@xm5F
\mathchardef\smallsmile="3\@xm60
\mathchardef\smallfrown="3\@xm61
\mathchardef\Subset="3\@xm62
\mathchardef\Supset="3\@xm63
\mathchardef\Cup="2\@xm64

\mathchardef\Cap="2\@xm65

\mathchardef\curlywedge="2\@xm66
\mathchardef\curlyvee="2\@xm67
\mathchardef\leftthreetimes="2\@xm68
\mathchardef\rightthreetimes="2\@xm69
\mathchardef\subseteqq="3\@xm6A
\mathchardef\supseteqq="3\@xm6B
\mathchardef\bumpeq="3\@xm6C
\mathchardef\Bumpeq="3\@xm6D
\mathchardef\lll="3\@xm6E

\mathchardef\ggg="3\@xm6F

\mathchardef\circledS="0\@xm73
\mathchardef\pitchfork="3\@xm74
\mathchardef\dotplus="2\@xm75
\mathchardef\backsim="3\@xm76
\mathchardef\backsimeq="3\@xm77
\mathchardef\complement="0\@xm7B
\mathchardef\intercal="2\@xm7C
\mathchardef\circledcirc="2\@xm7D
\mathchardef\circledast="2\@xm7E
\mathchardef\circleddash="2\@xm7F
\def\ulcorner{\delimiter"4\@xm70\@xm70 }
\def\urcorner{\delimiter"5\@xm71\@xm71 }
\def\llcorner{\delimiter"4\@xm78\@xm78 }
\def\lrcorner{\delimiter"5\@xm79\@xm79 }
\def\yen{\mathhexbox\@xm55 }
\def\checkmark{\mathhexbox\@xm58 }
\def\circledR{\mathhexbox\@xm72 }
\def\maltese{\mathhexbox\@xm7A }
\mathchardef\lvertneqq="3\@ym00
\mathchardef\gvertneqq="3\@ym01
\mathchardef\nleq="3\@ym02
\mathchardef\ngeq="3\@ym03
\mathchardef\nless="3\@ym04
\mathchardef\ngtr="3\@ym05
\mathchardef\nprec="3\@ym06
\mathchardef\nsucc="3\@ym07
\mathchardef\lneqq="3\@ym08
\mathchardef\gneqq="3\@ym09
\mathchardef\nleqslant="3\@ym0A
\mathchardef\ngeqslant="3\@ym0B
\mathchardef\lneq="3\@ym0C
\mathchardef\gneq="3\@ym0D
\mathchardef\npreceq="3\@ym0E
\mathchardef\nsucceq="3\@ym0F
\mathchardef\precnsim="3\@ym10
\mathchardef\succnsim="3\@ym11
\mathchardef\lnsim="3\@ym12
\mathchardef\gnsim="3\@ym13
\mathchardef\nleqq="3\@ym14
\mathchardef\ngeqq="3\@ym15
\mathchardef\precneqq="3\@ym16
\mathchardef\succneqq="3\@ym17
\mathchardef\precnapprox="3\@ym18
\mathchardef\succnapprox="3\@ym19
\mathchardef\lnapprox="3\@ym1A
\mathchardef\gnapprox="3\@ym1B
\mathchardef\nsim="3\@ym1C
\mathchardef\ncong="3\@ym1D

\mathchardef\varsubsetneq="3\@ym20
\mathchardef\varsupsetneq="3\@ym21
\mathchardef\nsubseteqq="3\@ym22
\mathchardef\nsupseteqq="3\@ym23
\mathchardef\subsetneqq="3\@ym24
\mathchardef\supsetneqq="3\@ym25
\mathchardef\varsubsetneqq="3\@ym26
\mathchardef\varsupsetneqq="3\@ym27
\mathchardef\subsetneq="3\@ym28
\mathchardef\supsetneq="3\@ym29
\mathchardef\nsubseteq="3\@ym2A
\mathchardef\nsupseteq="3\@ym2B
\mathchardef\nparallel="3\@ym2C
\mathchardef\nmid="3\@ym2D
\mathchardef\nshortmid="3\@ym2E
\mathchardef\nshortparallel="3\@ym2F
\mathchardef\nvdash="3\@ym30
\mathchardef\nVdash="3\@ym31
\mathchardef\nvDash="3\@ym32
\mathchardef\nVDash="3\@ym33
\mathchardef\ntrianglerighteq="3\@ym34
\mathchardef\ntrianglelefteq="3\@ym35
\mathchardef\ntriangleleft="3\@ym36
\mathchardef\ntriangleright="3\@ym37
\mathchardef\nleftarrow="3\@ym38
\mathchardef\nrightarrow="3\@ym39
\mathchardef\nLeftarrow="3\@ym3A
\mathchardef\nRightarrow="3\@ym3B
\mathchardef\nLeftrightarrow="3\@ym3C
\mathchardef\nleftrightarrow="3\@ym3D
\mathchardef\divideontimes="2\@ym3E
\mathchardef\varnothing="0\@ym3F
\mathchardef\nexists="0\@ym40
\mathchardef\mho="0\@ym66
\mathchardef\eth="0\@ym67
\mathchardef\eqsim="3\@ym68
\mathchardef\beth="0\@ym69
\mathchardef\gimel="0\@ym6A
\mathchardef\daleth="0\@ym6B
\mathchardef\lessdot="3\@ym6C
\mathchardef\gtrdot="3\@ym6D
\mathchardef\ltimes="2\@ym6E
\mathchardef\rtimes="2\@ym6F
\mathchardef\shortmid="3\@ym70
\mathchardef\shortparallel="3\@ym71
\mathchardef\smallsetminus="2\@ym72
\mathchardef\thicksim="3\@ym73
\mathchardef\thickapprox="3\@ym74
\mathchardef\approxeq="3\@ym75
\mathchardef\succapprox="3\@ym76
\mathchardef\precapprox="3\@ym77
\mathchardef\curvearrowleft="3\@ym78
\mathchardef\curvearrowright="3\@ym79
\mathchardef\digamma="0\@ym7A
\mathchardef\varkappa="0\@ym7B
\mathchardef\hslash="0\@ym7D
\mathchardef\hbar="0\@ym7E
\mathchardef\backepsilon="3\@ym7F


\def\Bbb{\ifmmode\let\next\Bbb@\else
\def\next{\errmessage{Use \string\Bbb\space only in math mode}}\fi\next}
\def\Bbb@#1{{\Bbb@@{#1}}}
\def\Bbb@@#1{\fam\ymfam#1}
\fi


\def\Nulle{0} 
\def\Afe{1}   
\def\Hae{2}   
\def\Hbe{3}   
\def\Hce{4}   
\def\Hde{5}   


\newcount\LastMac       \LastMac=\Nulle

\newskip\half      \half=5.5pt plus 1.5pt minus 2.25pt
\newskip\one       \one=11pt plus 3pt minus 5.5pt
\newskip\onehalf   \onehalf=16.5pt plus 5.5pt minus 8.25pt
\newskip\two       \two=22pt plus 5.5pt minus 11pt

\def\Half{\addvspace{\half}}
\def\One{\addvspace{\one}}
\def\OneHalf{\addvspace{\onehalf}}
\def\Two{\addvspace{\two}}


\def\Raggedright{
  \rightskip=\z@ plus \hsize\relax
}

\def\Fullout{
  \rightskip=\z@\relax
}

\def\Hang#1#2{
  \hangindent=#1%
  \hangafter=#2\relax
}


\newif\ifsp@page
\def\pagestyle#1{\csname ps@#1\endcsname}
\def\thispagestyle#1{\global\sp@pagetrue\gdef\sp@type{#1}}

\def\ps@titlepage{%
  \def\@oddhead{\eightpoint\noindent \the\CatchLine
    \ifprod@font\else\qquad Printed\ \today\fi \hfil}%
  \let\@evenhead=\@oddhead
}

\def\ps@headings{%
  \def\@oddhead{\elevenpoint\it\noindent
    \hfill\the\RightHeader\hskip1.5em\rm\folio}%
  \def\@evenhead{\elevenpoint\noindent
    \folio\hskip1.5em\it\the\LeftHeader\hfill}%
}

\def\ps@plate{%
  \def\@oddhead{\eightpoint\noindent\plt@cap\hfil}%
  \def\@evenhead{\eightpoint\noindent\plt@cap\hfil}%
}



\def\title#1{
  \bgroup
    \vbox to 8pt{\vss}%
    \seventeenpoint
    \Raggedright
    \noindent \strut{\bf #1}\par
  \egroup
}

\def\author#1{
  \bgroup
    \ifnum\LastMac=\Afe \OneHalf\else \vskip 21pt\fi
    \fourteenpoint
    \Raggedright
    \noindent \strut #1\par
    \vskip 3pt%
  \egroup
}

\def\affiliation#1{
  \bgroup
    \vskip -4pt%
    \eightpoint
    \Raggedright
    \noindent \strut {\it #1}\par
  \egroup
  \LastMac=\Afe\relax
}

\def\acceptedline#1{
  \bgroup
    \Two
    \eightpoint
    \Raggedright
    \noindent \strut #1\par
  \egroup
}

\long\def\abstract#1{%
  \bgroup
    \vskip 20pt%
    \everypar{\Hang{11pc}{0}}%
    \noindent{\ninebf ABSTRACT}\par
    \tenpoint
    \Fullout
    \noindent #1\par
  \egroup
}

\long\def\keywords#1{
  \bgroup
    \Half
    \everypar{\Hang{11pc}{0}}%
    \tenpoint
    \Fullout
    \noindent\hbox{\bf Key words:}\ #1\par
  \egroup
}


\def\maketitle{%
  \EndOpening
  \ifsinglecol \else \MakePage\fi
}



\def\Autonumber{
  \global\AutoNumbertrue  
}

\newif\ifAutoNumber \AutoNumberfalse
\newcount\Sec        
\newcount\SecSec
\newcount\SecSecSec

\Sec=\z@

\def\:{\let\@sptoken= } \:  
\def\:{\@xifnch} \expandafter\def\: {\futurelet\@tempc\@ifnch}

\def\@ifnextchar#1#2#3{%
  \let\@tempMACe #1%
  \def\@tempMACa{#2}%
  \def\@tempMACb{#3}%
  \futurelet \@tempMACc\@ifnch%
}

\def\@ifnch{%
\ifx \@tempMACc \@sptoken%
  \let\@tempMACd\@xifnch%
\else%
  \ifx \@tempMACc \@tempMACe%
    \let\@tempMACd\@tempMACa%
  \else%
    \let\@tempMACd\@tempMACb%
  \fi%
\fi%
\@tempMACd%
}

\def\@ifstar#1#2{\@ifnextchar *{\def\@tempMACa*{#1}\@tempMACa}{#2}}

\newskip\@tempskipb

\def\addvspace#1{%
  \ifvmode\else \endgraf\fi%
  \ifdim\lastskip=\z@%
    \vskip #1\relax%
  \else%
    \@tempskipb#1\relax\@xaddvskip%
  \fi%
}

\def\@xaddvskip{%
  \ifdim\lastskip<\@tempskipb%
    \vskip-\lastskip%
    \vskip\@tempskipb\relax%
  \else%
    \ifdim\@tempskipb<\z@%
      \ifdim\lastskip<\z@ \else%
        \advance\@tempskipb\lastskip%
        \vskip-\lastskip\vskip\@tempskipb%
      \fi%
    \fi%
  \fi%
}

\newskip\@tmpSKIP

\def\addpen#1{%
  \ifvmode
    \if@nobreak
    \else
      \ifdim\lastskip=\z@
        \penalty#1\relax
      \else
        \@tmpSKIP=\lastskip
        \vskip -\lastskip
        \penalty#1\vskip\@tmpSKIP
      \fi
    \fi
  \fi
}

\newcount\@clubpen   \@clubpen=\clubpenalty
\newif\if@nobreak    \@nobreakfalse

\def\@noafterindent{%
  \global\@nobreaktrue
  \everypar{\if@nobreak
              \global\@nobreakfalse
              \clubpenalty \@M
              {\setbox\z@\lastbox}%
              \LastMac=\Nulle\relax%
            \else
              \clubpenalty \@clubpen
              \everypar{}%
            \fi}
}

\newcount\gds@cbrk   \gds@cbrk=-300

\def\@nohdbrk{\interlinepenalty \@M\relax}

\let\@par=\par
\def\@restorepar{\def\par{\@par}}

\newif\if@endpe   \@endpefalse
 
\def\@doendpe{\@endpetrue \@nobreakfalse \LastMac=\Nulle\relax%
     \def\par{\@restorepar\everypar{}\par\@endpefalse}%
              \everypar{\setbox\z@\lastbox\everypar{}\@endpefalse}%
}

\def\section{\@ifstar{\@ssection}{\@section}}

\def\@section#1{
  \if@nobreak
    \everypar{}%
    \ifnum\LastMac=\Hae \addvspace{\half}\fi
  \else
    \addpen{\gds@cbrk}%
    \addvspace{\two}%
  \fi
  \bgroup
    \ninepoint\bf
    \Raggedright
    \ifAutoNumber
      \global\advance\Sec \@ne
      \noindent\@nohdbrk\number\Sec\hskip 1pc \uppercase{#1}\par
      \global\SecSec=\z@
    \else
      \noindent\@nohdbrk\uppercase{#1}\par
    \fi
  \egroup
  \nobreak
  \vskip\half
  \nobreak
  \@noafterindent
  \LastMac=\Hae\relax
}

\def\@ssection#1{
  \if@nobreak
    \everypar{}%
    \ifnum\LastMac=\Hae \addvspace{\half}\fi
  \else
    \addpen{\gds@cbrk}%
    \addvspace{\two}%
  \fi
  \bgroup
    \ninepoint\bf
    \Raggedright
    \noindent\@nohdbrk\uppercase{#1}\par
  \egroup
  \nobreak
  \vskip\half
  \nobreak
  \@noafterindent
  \LastMac=\Hae\relax
}

\def\subsection#1{
  \if@nobreak
    \everypar{}%
    \ifnum\LastMac=\Hae \addvspace{1pt plus 1pt minus .5pt}\fi
  \else
    \addpen{\gds@cbrk}%
    \addvspace{\onehalf}%
  \fi
  \bgroup
    \ninepoint\bf
    \Raggedright
    \ifAutoNumber
      \global\advance\SecSec \@ne
      \noindent\@nohdbrk\number\Sec.\number\SecSec \hskip 1pc\relax #1\par
      \global\SecSecSec=\z@
    \else
      \noindent\@nohdbrk #1\par
    \fi
  \egroup
  \nobreak
  \vskip\half
  \nobreak
  \@noafterindent
  \LastMac=\Hbe\relax
}

\def\subsubsection#1{
  \if@nobreak
    \everypar{}%
    \ifnum\LastMac=\Hbe \addvspace{1pt plus 1pt minus .5pt}\fi
  \else
    \addpen{\gds@cbrk}%
    \addvspace{\onehalf}%
  \fi
  \bgroup
    \ninepoint\it
    \Raggedright
    \ifAutoNumber
      \global\advance\SecSecSec \@ne
      \noindent\@nohdbrk\number\Sec.\number\SecSec.\number\SecSecSec
        \hskip 1pc\relax #1\par
    \else
      \noindent\@nohdbrk #1\par
    \fi
  \egroup
  \nobreak
  \vskip\half
  \nobreak
  \@noafterindent
  \LastMac=\Hce\relax
}

\def\paragraph#1{
  \if@nobreak
    \everypar{}%
  \else
    \addpen{\gds@cbrk}%
    \addvspace{\one}%
  \fi%
  \bgroup%
    \ninepoint\it
    \noindent #1\ \nobreak%
  \egroup
  \LastMac=\Hde\relax
  \ignorespaces
}




\def\beginlist{%
  \par\if@nobreak \else\addvspace{\half}\fi%
  \bgroup%
    \ninepoint
    \let\item=\list@item%
}

\def\list@item{%
  \par\noindent\hskip 1em\relax%
  \ignorespaces%
}

\def\endlist{\par\egroup\addvspace{\half}\@doendpe}


\def\beginrefs{%
  \par
  \bgroup
    \eightpoint
    \Raggedright
    \let\bibitem=\bib@item
}

\def\bib@item{%
  \par\parindent=1.5em\Hang{1.5em}{1}%
  \everypar={\Hang{1.5em}{1}\ignorespaces}%
  \noindent\ignorespaces
}

\def\endrefs{\par\egroup\@doendpe}


\newtoks\CatchLine

\def\@journal{Mon.\ Not.\ R.\ Astron.\ Soc.\ }  
\def\@pubyear{1996}        
\def\@pagerange{000--000}  
\def\@volume{000}          
\def\@microfiche{}         %

\def\pubyear#1{\gdef\@pubyear{#1}\@makecatchline}
\def\pagerange#1{\gdef\@pagerange{#1}\@makecatchline}
\def\volume#1{\gdef\@volume{#1}\@makecatchline}
\def\microfiche#1{\gdef\@microfiche{and Microfiche\ #1}\@makecatchline}

\def\@makecatchline{%
  \global\CatchLine{%
    {\rm \@journal {\bf \@volume},\ \@pagerange\ (\@pubyear)\ \@microfiche}}%
}

\@makecatchline 

\newtoks\LeftHeader
\def\shortauthor#1{
  \global\LeftHeader{#1}%
}

\newtoks\RightHeader
\def\shorttitle#1{
  \global\RightHeader{#1}%
}

\def\PageHead{
  \begingroup
    \ifsp@page
      \csname ps@\sp@type\endcsname
      \global\sp@pagefalse
    \fi
    \ifodd\pageno
      \let\the@head=\@oddhead
    \else
      \let\the@head=\@evenhead
    \fi
    \vbox to \z@{\vskip-22.5\p@%
      \hbox to \PageWidth{\vbox to8.5\p@{}%
        \the@head
      }%
    \vss}%
  \endgroup
  \nointerlineskip
}

\def\today{%
  \number\day\space
  \ifcase\month\or January\or February\or March\or April\or May\or June\or
    July\or August\or September\or October\or November\or December\fi
  \space\number\year%
}

\def\PageFoot{} 

\def\authorcomment#1{%
  \gdef\PageFoot{%
    \nointerlineskip%
    \vbox to 22pt{\vfil%
      \hbox to \PageWidth{\elevenpoint\noindent \hfil #1 \hfil}}%
  }%
}


\newif\ifplate@page
\newbox\plt@box

\def\beginplatepage{%
  \let\plate=\plate@head
  \let\caption=\fig@caption
  \global\setbox\plt@box=\vbox\bgroup
  \TEMPDIMEN=\PageWidth 
  \hsize=\PageWidth\relax
}

\def\endplatepage{\par\egroup\global\plate@pagetrue}
\def\plate@head#1{\gdef\plt@cap{#1}}


\def\letters{%
  \gdef\folio{\ifnum\pageno<\z@ L\romannumeral-\pageno
    \else L\number\pageno \fi}%
}


\everydisplay{\displaysetup}

\newif\ifeqno
\newif\ifleqno

\def\displaysetup#1$${%
 \displaytest#1\eqno\eqno\displaytest
}

\def\displaytest#1\eqno#2\eqno#3\displaytest{%
 \if!#3!\ldisplaytest#1\leqno\leqno\ldisplaytest
 \else\eqnotrue\leqnofalse\def\eqn{#2}\def\eq{#1}\fi
 \generaldisplay$$}

\def\ldisplaytest#1\leqno#2\leqno#3\ldisplaytest{%
 \def\eq{#1}%
 \if!#3!\eqnofalse\else\eqnotrue\leqnotrue
  \def\eqn{#2}\fi}

\def\generaldisplay{%
\ifeqno \ifleqno 
   \hbox to \hsize{\noindent
     $\displaystyle\eq$\hfil$\displaystyle\eqn$}
  \else
    \hbox to \hsize{\noindent
     $\displaystyle\eq$\hfil$\displaystyle\eqn$}
  \fi
 \else
 \hbox to \hsize{\vbox{\noindent
  $\displaystyle\eq$\hfil}}
 \fi
}


\def\@notice{%
  \par\Two%
  \noindent{\b@ls{11pt}\ninerm This paper has been produced using the
    Blackwell Scientific Publications \TeX\ macros.\par}%
}

\outer\def\bye{\@notice\par\vfill\supereject\end}


\def\start@mess{%
  Monthly notices of the RAS journal style (\@typeface)\space
    v\@version,\space \@verdate.%
}

\everyjob{\Warn{\start@mess}}



\newif\if@debug \@debugfalse  

\def\Print#1{\if@debug\immediate\write16{#1}\else \fi}
\def\Warn#1{\immediate\write16{#1}}
\def\wlog#1{}

\newcount\Iteration 

\def\Single{0} \def\Double{1}                 
\def\Figure{0} \def\Table{1}                  

\def\InStack{0}  
\def\InZoneA{1}
\def\InZoneB{2}
\def\InZoneC{3}

\newcount\TEMPCOUNT 
\newdimen\TEMPDIMEN 
\newbox\TEMPBOX     
\newbox\VOIDBOX     

\newcount\LengthOfStack 
\newcount\MaxItems      
\newcount\StackPointer
\newcount\Point         
\newcount\NextFigure    
\newcount\NextTable     
\newcount\NextItem      

\newcount\StatusStack   
\newcount\NumStack      
\newcount\TypeStack     
\newcount\SpanStack     
\newcount\BoxStack      

\newcount\ItemSTATUS    
\newcount\ItemNUMBER    
\newcount\ItemTYPE      
\newcount\ItemSPAN      
\newbox\ItemBOX         
\newdimen\ItemSIZE      

\newdimen\PageHeight    
\newdimen\TextLeading   
\newdimen\Feathering    
\newcount\LinesPerPage  
\newdimen\ColumnWidth   
\newdimen\ColumnGap     
\newdimen\PageWidth     
\newdimen\BodgeHeight   
\newcount\Leading       

\newdimen\ZoneBSize  
\newdimen\TextSize   
\newbox\ZoneABOX     
\newbox\ZoneBBOX     
\newbox\ZoneCBOX     

\newif\ifFirstSingleItem
\newif\ifFirstZoneA
\newif\ifMakePageInComplete
\newif\ifMoreFigures \MoreFiguresfalse 
\newif\ifMoreTables  \MoreTablesfalse  

\newif\ifFigInZoneB 
\newif\ifFigInZoneC 
\newif\ifTabInZoneB 
\newif\ifTabInZoneC

\newif\ifZoneAFullPage

\newbox\MidBOX    
\newbox\LeftBOX
\newbox\RightBOX
\newbox\PageBOX   

\newif\ifLeftCOL  
\LeftCOLtrue

\newdimen\ZoneBAdjust

\newcount\ItemFits
\def\Yes{1}
\def\No{2}


\MaxItems=15
\NextFigure=\z@        
\NextTable=\@ne

\BodgeHeight=6pt
\TextLeading=11pt    
\Leading=11
\Feathering=\z@      
\LinesPerPage=61     
\topskip=\TextLeading
\ColumnWidth=20pc    
\ColumnGap=2pc       

\newskip\ItemSepamount  
\ItemSepamount=\TextLeading plus \TextLeading minus 4pt

\parskip=\z@ plus .1pt
\parindent=18pt
\widowpenalty=\z@
\clubpenalty=10000
\tolerance=1500
\hbadness=1500
\abovedisplayskip=6pt plus 2pt minus 2pt
\belowdisplayskip=6pt plus 2pt minus 2pt
\abovedisplayshortskip=6pt plus 2pt minus 2pt
\belowdisplayshortskip=6pt plus 2pt minus 2pt

\ninepoint 


\PageHeight=682pt

\PageWidth=2\ColumnWidth
\advance\PageWidth by \ColumnGap

\pagestyle{headings}




\newcount\DUMMY \StatusStack=\allocationnumber
\newcount\DUMMY \newcount\DUMMY \newcount\DUMMY 
\newcount\DUMMY \newcount\DUMMY \newcount\DUMMY 
\newcount\DUMMY \newcount\DUMMY \newcount\DUMMY
\newcount\DUMMY \newcount\DUMMY \newcount\DUMMY 
\newcount\DUMMY \newcount\DUMMY \newcount\DUMMY

\newcount\DUMMY \NumStack=\allocationnumber
\newcount\DUMMY \newcount\DUMMY \newcount\DUMMY 
\newcount\DUMMY \newcount\DUMMY \newcount\DUMMY 
\newcount\DUMMY \newcount\DUMMY \newcount\DUMMY 
\newcount\DUMMY \newcount\DUMMY \newcount\DUMMY 
\newcount\DUMMY \newcount\DUMMY \newcount\DUMMY

\newcount\DUMMY \TypeStack=\allocationnumber
\newcount\DUMMY \newcount\DUMMY \newcount\DUMMY 
\newcount\DUMMY \newcount\DUMMY \newcount\DUMMY 
\newcount\DUMMY \newcount\DUMMY \newcount\DUMMY 
\newcount\DUMMY \newcount\DUMMY \newcount\DUMMY 
\newcount\DUMMY \newcount\DUMMY \newcount\DUMMY

\newcount\DUMMY \SpanStack=\allocationnumber
\newcount\DUMMY \newcount\DUMMY \newcount\DUMMY 
\newcount\DUMMY \newcount\DUMMY \newcount\DUMMY 
\newcount\DUMMY \newcount\DUMMY \newcount\DUMMY 
\newcount\DUMMY \newcount\DUMMY \newcount\DUMMY 
\newcount\DUMMY \newcount\DUMMY \newcount\DUMMY

\newbox\DUMMY   \BoxStack=\allocationnumber
\newbox\DUMMY   \newbox\DUMMY \newbox\DUMMY 
\newbox\DUMMY   \newbox\DUMMY \newbox\DUMMY 
\newbox\DUMMY   \newbox\DUMMY \newbox\DUMMY 
\newbox\DUMMY   \newbox\DUMMY \newbox\DUMMY 
\newbox\DUMMY   \newbox\DUMMY \newbox\DUMMY

\def\wlog{\immediate\write\m@ne}


\def\GetItemAll#1{%
 \GetItemSTATUS{#1}
 \GetItemNUMBER{#1}
 \GetItemTYPE{#1}
 \GetItemSPAN{#1}
 \GetItemBOX{#1}
}

\def\GetItemSTATUS#1{%
 \Point=\StatusStack
 \advance\Point by #1
 \global\ItemSTATUS=\count\Point
}

\def\GetItemNUMBER#1{%
 \Point=\NumStack
 \advance\Point by #1
 \global\ItemNUMBER=\count\Point
}

\def\GetItemTYPE#1{%
 \Point=\TypeStack
 \advance\Point by #1
 \global\ItemTYPE=\count\Point
}

\def\GetItemSPAN#1{%
 \Point\SpanStack
 \advance\Point by #1
 \global\ItemSPAN=\count\Point
}

\def\GetItemBOX#1{%
 \Point=\BoxStack
 \advance\Point by #1
 \global\setbox\ItemBOX=\vbox{\copy\Point}
 \global\ItemSIZE=\ht\ItemBOX
 \global\advance\ItemSIZE by \dp\ItemBOX
 \TEMPCOUNT=\ItemSIZE
 \divide\TEMPCOUNT by \Leading
 \divide\TEMPCOUNT by 65536
 \advance\TEMPCOUNT \@ne
 \ItemSIZE=\TEMPCOUNT pt
 \global\multiply\ItemSIZE by \Leading
}


\def\JoinStack{%
 \ifnum\LengthOfStack=\MaxItems 
  \Warn{WARNING: Stack is full...some items will be lost!}
 \else
  \Point=\StatusStack
  \advance\Point by \LengthOfStack
  \global\count\Point=\ItemSTATUS
  \Point=\NumStack
  \advance\Point by \LengthOfStack
  \global\count\Point=\ItemNUMBER
  \Point=\TypeStack
  \advance\Point by \LengthOfStack
  \global\count\Point=\ItemTYPE
  \Point\SpanStack
  \advance\Point by \LengthOfStack
  \global\count\Point=\ItemSPAN
  \Point=\BoxStack
  \advance\Point by \LengthOfStack
  \global\setbox\Point=\vbox{\copy\ItemBOX}
  \global\advance\LengthOfStack \@ne
  \ifnum\ItemTYPE=\Figure 
   \global\MoreFigurestrue
  \else
   \global\MoreTablestrue
  \fi
 \fi
}


\def\LeaveStack#1{%
 {\Iteration=#1
 \loop
 \ifnum\Iteration<\LengthOfStack
  \advance\Iteration \@ne
  \GetItemSTATUS{\Iteration}
   \advance\Point by \m@ne
   \global\count\Point=\ItemSTATUS
  \GetItemNUMBER{\Iteration}
   \advance\Point by \m@ne
   \global\count\Point=\ItemNUMBER
  \GetItemTYPE{\Iteration}
   \advance\Point by \m@ne
   \global\count\Point=\ItemTYPE
  \GetItemSPAN{\Iteration}
   \advance\Point by \m@ne
   \global\count\Point=\ItemSPAN
  \GetItemBOX{\Iteration}
   \advance\Point by \m@ne
   \global\setbox\Point=\vbox{\copy\ItemBOX}
 \repeat}
 \global\advance\LengthOfStack by \m@ne
}


\newif\ifStackNotClean

\def\CleanStack{%
 \StackNotCleantrue
 {\Iteration=\z@
  \loop
   \ifStackNotClean
    \GetItemSTATUS{\Iteration}
    \ifnum\ItemSTATUS=\InStack
     \advance\Iteration \@ne
     \else
      \LeaveStack{\Iteration}
    \fi
   \ifnum\LengthOfStack<\Iteration
    \StackNotCleanfalse
   \fi
 \repeat}
}


\def\FindItem#1#2{%
 \global\StackPointer=\m@ne 
 {\Iteration=\z@
  \loop
  \ifnum\Iteration<\LengthOfStack
   \GetItemSTATUS{\Iteration}
   \ifnum\ItemSTATUS=\InStack
    \GetItemTYPE{\Iteration}
    \ifnum\ItemTYPE=#1
     \GetItemNUMBER{\Iteration}
     \ifnum\ItemNUMBER=#2
      \global\StackPointer=\Iteration
      \Iteration=\LengthOfStack 
     \fi
    \fi
   \fi
  \advance\Iteration \@ne
 \repeat}
}


\def\FindNext{%
 \global\StackPointer=\m@ne 
 {\Iteration=\z@
  \loop
  \ifnum\Iteration<\LengthOfStack
   \GetItemSTATUS{\Iteration}
   \ifnum\ItemSTATUS=\InStack
    \GetItemTYPE{\Iteration}
   \ifnum\ItemTYPE=\Figure
    \ifMoreFigures
      \global\NextItem=\Figure
      \global\StackPointer=\Iteration
      \Iteration=\LengthOfStack 
    \fi
   \fi
   \ifnum\ItemTYPE=\Table
    \ifMoreTables
      \global\NextItem=\Table
      \global\StackPointer=\Iteration
      \Iteration=\LengthOfStack 
    \fi
   \fi
  \fi
  \advance\Iteration \@ne
 \repeat}
}


\def\ChangeStatus#1#2{%
 \Point=\StatusStack
 \advance\Point by #1
 \global\count\Point=#2
}



\def\Zone{\InZoneA}

\ZoneBAdjust=\z@

\def\MakePage{
 \global\ZoneBSize=\PageHeight
 \global\TextSize=\ZoneBSize
 \global\ZoneAFullPagefalse
 \global\topskip=\TextLeading
 \MakePageInCompletetrue
 \MoreFigurestrue
 \MoreTablestrue
 \FigInZoneBfalse
 \FigInZoneCfalse
 \TabInZoneBfalse
 \TabInZoneCfalse
 \global\FirstSingleItemtrue
 \global\FirstZoneAtrue
 \global\setbox\ZoneABOX=\box\VOIDBOX
 \global\setbox\ZoneBBOX=\box\VOIDBOX
 \global\setbox\ZoneCBOX=\box\VOIDBOX
 \loop
  \ifMakePageInComplete
 \FindNext
 \ifnum\StackPointer=\m@ne
  \NextItem=\m@ne
  \MoreFiguresfalse
  \MoreTablesfalse
 \fi
 \ifnum\NextItem=\Figure
   \FindItem{\Figure}{\NextFigure}
   \ifnum\StackPointer=\m@ne \global\MoreFiguresfalse
   \else
    \GetItemSPAN{\StackPointer}
    \ifnum\ItemSPAN=\Single \def\Zone{\InZoneB}\relax
     \ifFigInZoneC \global\MoreFiguresfalse\fi
    \else
     \def\Zone{\InZoneA}
     \ifFigInZoneB \def\Zone{\InZoneC}\fi
    \fi
   \fi
   \ifMoreFigures\Print{}\FigureItems\fi
 \fi
\ifnum\NextItem=\Table
   \FindItem{\Table}{\NextTable}
   \ifnum\StackPointer=\m@ne \global\MoreTablesfalse
   \else
    \GetItemSPAN{\StackPointer}
    \ifnum\ItemSPAN=\Single\relax
     \ifTabInZoneC \global\MoreTablesfalse\fi
    \else
     \def\Zone{\InZoneA}
     \ifTabInZoneB \def\Zone{\InZoneC}\fi
    \fi
   \fi
   \ifMoreTables\Print{}\TableItems\fi
 \fi
   \MakePageInCompletefalse 
   \ifMoreFigures\MakePageInCompletetrue\fi
   \ifMoreTables\MakePageInCompletetrue\fi
 \repeat
 \ifZoneAFullPage
  \global\TextSize=\z@
  \global\ZoneBSize=\z@
  \global\vsize=\z@\relax
  \global\topskip=\z@\relax
  \vbox to \z@{\vss}
  \eject
 \else
 \global\advance\ZoneBSize by -\ZoneBAdjust
 \global\vsize=\ZoneBSize
 \global\hsize=\ColumnWidth
 \global\ZoneBAdjust=\z@
 \ifdim\TextSize<23pt
 \Warn{}
 \Warn{* Making column fall short: TextSize=\the\TextSize *}
 \vskip-\lastskip\eject\fi
 \fi
}

\def\MakeRightCol{
 \global\TextSize=\ZoneBSize
 \MakePageInCompletetrue
 \MoreFigurestrue
 \MoreTablestrue
 \global\FirstSingleItemtrue
 \global\setbox\ZoneBBOX=\box\VOIDBOX
 \def\Zone{\InZoneB}
 \loop
  \ifMakePageInComplete
 \FindNext
 \ifnum\StackPointer=\m@ne
  \NextItem=\m@ne
  \MoreFiguresfalse
  \MoreTablesfalse
 \fi
 \ifnum\NextItem=\Figure
   \FindItem{\Figure}{\NextFigure}
   \ifnum\StackPointer=\m@ne \MoreFiguresfalse
   \else
    \GetItemSPAN{\StackPointer}
    \ifnum\ItemSPAN=\Double\relax
     \MoreFiguresfalse\fi
   \fi
   \ifMoreFigures\Print{}\FigureItems\fi
 \fi
 \ifnum\NextItem=\Table
   \FindItem{\Table}{\NextTable}
   \ifnum\StackPointer=\m@ne \MoreTablesfalse
   \else
    \GetItemSPAN{\StackPointer}
    \ifnum\ItemSPAN=\Double\relax
     \MoreTablesfalse\fi
   \fi
   \ifMoreTables\Print{}\TableItems\fi
 \fi
   \MakePageInCompletefalse 
   \ifMoreFigures\MakePageInCompletetrue\fi
   \ifMoreTables\MakePageInCompletetrue\fi
 \repeat
 \ifZoneAFullPage
  \global\TextSize=\z@
  \global\ZoneBSize=\z@
  \global\vsize=\z@\relax
  \global\topskip=\z@\relax
  \vbox to \z@{\vss}
  \eject
 \else
 \global\vsize=\ZoneBSize
 \global\hsize=\ColumnWidth
 \ifdim\TextSize<23pt
 \Warn{}
 \Warn{* Making column fall short: TextSize=\the\TextSize *}
 \vskip-\lastskip\eject\fi
\fi
}

\def\FigureItems{
 \Print{Considering...}
 \ShowItem{\StackPointer}
 \GetItemBOX{\StackPointer} 
 \GetItemSPAN{\StackPointer}
  \CheckFitInZone 
  \ifnum\ItemFits=\Yes
   \ifnum\ItemSPAN=\Single
     \ChangeStatus{\StackPointer}{\InZoneB} 
     \global\FigInZoneBtrue
     \ifFirstSingleItem
      \hbox{}\vskip-\BodgeHeight
     \global\advance\ItemSIZE by \TextLeading
     \fi
     \unvbox\ItemBOX\ItemSep
     \global\FirstSingleItemfalse
     \global\advance\TextSize by -\ItemSIZE
     \global\advance\TextSize by -\TextLeading
   \else
    \ifFirstZoneA
     \global\advance\ItemSIZE by \TextLeading
     \global\FirstZoneAfalse\fi
    \global\advance\TextSize by -\ItemSIZE
    \global\advance\TextSize by -\TextLeading
    \global\advance\ZoneBSize by -\ItemSIZE
    \global\advance\ZoneBSize by -\TextLeading
    \ifFigInZoneB\relax
     \else
     \ifdim\TextSize<3\TextLeading
     \global\ZoneAFullPagetrue
     \fi
    \fi
    \ChangeStatus{\StackPointer}{\Zone}
    \ifnum\Zone=\InZoneC \global\FigInZoneCtrue\fi
  \fi
   \Print{TextSize=\the\TextSize}
   \Print{ZoneBSize=\the\ZoneBSize}
  \global\advance\NextFigure \@ne
   \Print{This figure has been placed.}
  \else
   \Print{No space available for this figure...holding over.}
   \Print{}
   \global\MoreFiguresfalse
  \fi
}

\def\TableItems{
 \Print{Considering...}
 \ShowItem{\StackPointer}
 \GetItemBOX{\StackPointer} 
 \GetItemSPAN{\StackPointer}
  \CheckFitInZone 
  \ifnum\ItemFits=\Yes
   \ifnum\ItemSPAN=\Single
    \ChangeStatus{\StackPointer}{\InZoneB}
     \global\TabInZoneBtrue
     \ifFirstSingleItem
      \hbox{}\vskip-\BodgeHeight
     \global\advance\ItemSIZE by \TextLeading
     \fi
     \unvbox\ItemBOX\ItemSep
     \global\FirstSingleItemfalse
     \global\advance\TextSize by -\ItemSIZE
     \global\advance\TextSize by -\TextLeading
   \else
    \ifFirstZoneA
    \global\advance\ItemSIZE by \TextLeading
    \global\FirstZoneAfalse\fi
    \global\advance\TextSize by -\ItemSIZE
    \global\advance\TextSize by -\TextLeading
    \global\advance\ZoneBSize by -\ItemSIZE
    \global\advance\ZoneBSize by -\TextLeading
    \ifFigInZoneB\relax
     \else
     \ifdim\TextSize<3\TextLeading
     \global\ZoneAFullPagetrue
     \fi
    \fi
    \ChangeStatus{\StackPointer}{\Zone}
    \ifnum\Zone=\InZoneC \global\TabInZoneCtrue\fi
   \fi
  \global\advance\NextTable \@ne
   \Print{This table has been placed.}
  \else
  \Print{No space available for this table...holding over.}
   \Print{}
   \global\MoreTablesfalse
  \fi
}


\def\CheckFitInZone{%
{\advance\TextSize by -\ItemSIZE
 \advance\TextSize by -\TextLeading
 \ifFirstSingleItem
  \advance\TextSize by \TextLeading
 \fi
 \ifnum\Zone=\InZoneA\relax
  \else \advance\TextSize by -\ZoneBAdjust
 \fi
 \ifdim\TextSize<3\TextLeading \global\ItemFits=\No
 \else \global\ItemFits=\Yes\fi}
}

\def\BeginOpening{%
  \thispagestyle{titlepage}%
  \global\setbox\ItemBOX=\vbox\bgroup%
    \hsize=\PageWidth%
    \hrule height \z@
    \ifsinglecol\vskip 6pt\fi 
}

\let\begintopmatter=\BeginOpening  

\def\EndOpening{%
  \One
  \egroup
  \ifsinglecol
    \box\ItemBOX%
    \vskip\TextLeading plus 2\TextLeading
    \@noafterindent
  \else
    \ItemNUMBER=\z@%
    \ItemTYPE=\Figure
    \ItemSPAN=\Double
    \ItemSTATUS=\InStack
    \JoinStack
  \fi
}


\newif\if@here  \@herefalse

\def\no@float{\global\@heretrue}
\let\nofloat=\relax 

\def\beginfigure{%
  \@ifstar{\global\@dfloattrue \@bfigure}{\global\@dfloatfalse \@bfigure}%
}

\def\@bfigure#1{%
  \par
  \if@dfloat
    \ItemSPAN=\Double
    \TEMPDIMEN=\PageWidth
  \else
    \ItemSPAN=\Single
    \TEMPDIMEN=\ColumnWidth
  \fi
  \ifsinglecol
    \TEMPDIMEN=\PageWidth
  \else
    \ItemSTATUS=\InStack
    \ItemNUMBER=#1%
    \ItemTYPE=\Figure
  \fi
  \bgroup
    \hsize=\TEMPDIMEN
    \global\setbox\ItemBOX=\vbox\bgroup
      \eightpoint\nostb@ls{10pt}%
      \let\caption=\fig@caption
      \ifsinglecol \let\nofloat=\no@float\fi
}

\def\fig@caption#1{%
  \vskip 5.5pt plus 6pt%
  \bgroup 
    \eightpoint\nostb@ls{10pt}%
    \setbox\TEMPBOX=\hbox{#1}%
    \ifdim\wd\TEMPBOX>\TEMPDIMEN
      \noindent \unhbox\TEMPBOX\par
    \else
      \hbox to \hsize{\hfil\unhbox\TEMPBOX\hfil}%
    \fi
  \egroup
}

\def\endfigure{%
  \par\egroup 
  \egroup
  \ifsinglecol
    \if@here \midinsert\global\@herefalse\else \topinsert\fi
      \unvbox\ItemBOX
    \endinsert
  \else
    \JoinStack
    \Print{Processing source for figure \the\ItemNUMBER}%
  \fi
}


\newbox\tab@cap@box
\def\tab@caption#1{\global\setbox\tab@cap@box=\hbox{#1\par}}

\newtoks\tab@txt@toks
\long\def\tab@txt#1{\global\tab@txt@toks={#1}\global\table@txttrue}

\newif\iftable@txt  \table@txtfalse
\newif\if@dfloat    \@dfloatfalse

\def\begintable{%
  \@ifstar{\global\@dfloattrue \@btable}{\global\@dfloatfalse \@btable}%
}

\def\@btable#1{%
  \par
  \if@dfloat
    \ItemSPAN=\Double
    \TEMPDIMEN=\PageWidth
  \else
    \ItemSPAN=\Single
    \TEMPDIMEN=\ColumnWidth
  \fi
  \ifsinglecol
    \TEMPDIMEN=\PageWidth
  \else
    \ItemSTATUS=\InStack
    \ItemNUMBER=#1%
    \ItemTYPE=\Table
  \fi
  \bgroup
    \eightpoint\nostb@ls{10pt}%
    \global\setbox\ItemBOX=\vbox\bgroup
      \let\caption=\tab@caption
      \let\tabletext=\tab@txt
      \ifsinglecol \let\nofloat=\no@float\fi
}

\def\endtable{%
  \par\egroup 
  \egroup
  \setbox\TEMPBOX=\hbox to \TEMPDIMEN{%
    \hss
    \vbox{%
      \hsize=\wd\ItemBOX
      \ifvoid\tab@cap@box
      \else
        \noindent\unhbox\tab@cap@box
        \vskip 5.5pt plus 6pt%
      \fi
      \box\ItemBOX
      \iftable@txt
        \vskip 10pt%
        \eightpoint\nostb@ls{10pt}%
        \noindent\the\tab@txt@toks
        \global\table@txtfalse
      \fi
    }%
    \hss
  }%
  \ifsinglecol
    \if@here \midinsert\global\@herefalse\else \topinsert\fi
      \box\TEMPBOX
    \endinsert
  \else
    \global\setbox\ItemBOX=\box\TEMPBOX
    \JoinStack
    \Print{Processing source for table \the\ItemNUMBER}%
  \fi
}

\def\UnloadZoneA{%
\FirstZoneAtrue
 \Iteration=\z@
  \loop
   \ifnum\Iteration<\LengthOfStack
    \GetItemSTATUS{\Iteration}
    \ifnum\ItemSTATUS=\InZoneA
     \GetItemBOX{\Iteration}
     \ifFirstZoneA \vbox to \BodgeHeight{\vfil}%
     \FirstZoneAfalse\fi
     \unvbox\ItemBOX\ItemSep
     \LeaveStack{\Iteration}
     \else
     \advance\Iteration \@ne
   \fi
 \repeat
}

\def\UnloadZoneC{%
\Iteration=\z@
  \loop
   \ifnum\Iteration<\LengthOfStack
    \GetItemSTATUS{\Iteration}
    \ifnum\ItemSTATUS=\InZoneC
     \GetItemBOX{\Iteration}
     \ItemSep\unvbox\ItemBOX
     \LeaveStack{\Iteration}
     \else
     \advance\Iteration \@ne
   \fi
 \repeat
}


\def\ShowItem#1{
  {\GetItemAll{#1}
  \Print{\the#1:
  {TYPE=\ifnum\ItemTYPE=\Figure Figure\else Table\fi}
  {NUMBER=\the\ItemNUMBER}
  {SPAN=\ifnum\ItemSPAN=\Single Single\else Double\fi}
  {SIZE=\the\ItemSIZE}}}
}

\def\ShowStack{%
 \Print{}
 \Print{LengthOfStack = \the\LengthOfStack}
 \ifnum\LengthOfStack=\z@ \Print{Stack is empty}\fi
 \Iteration=\z@
 \loop
 \ifnum\Iteration<\LengthOfStack
  \ShowItem{\Iteration}
  \advance\Iteration \@ne
 \repeat
}

\def\B#1#2{%
\hbox{\vrule\kern-0.4pt\vbox to #2{%
\hrule width #1\vfill\hrule}\kern-0.4pt\vrule}
}


\newif\ifsinglecol   \singlecolfalse

\def\onecolumn{%
  \global\output={\singlecoloutput}%
  \global\hsize=\PageWidth
  \global\vsize=\PageHeight
  \global\ColumnWidth=\hsize
  \global\TextLeading=12pt
  \global\Leading=12
  \global\singlecoltrue
  \global\let\onecolumn=\relax
  \global\let\footnote=\sing@footnote
  \global\let\vfootnote=\sing@vfootnote
  \ninepoint 
  \message{(Single column)}%
}

\def\singlecoloutput{%
  \shipout\vbox{\PageHead\pagebody\PageFoot}%
  \advancepageno
  \ifplate@page
    \shipout\vbox{%
      \sp@pagetrue
      \def\sp@type{plate}%
      \global\plate@pagefalse
      \PageHead\vbox to \PageHeight{\unvbox\plt@box\vfil}\PageFoot%
    }%
    \message{[plate]}%
    \advancepageno
  \fi
  \ifnum\outputpenalty>-\@MM \else\dosupereject\fi%
}

\def\ItemSep{\vskip\ItemSepamount\relax}

\def\ItemSepbreak{\par\ifdim\lastskip<\ItemSepamount
  \removelastskip\penalty-200\ItemSep\fi%
}


\let\@@endinsert=\endinsert 

\def\endinsert{\egroup 
  \if@mid \dimen@\ht\z@ \advance\dimen@\dp\z@ \advance\dimen@12\p@
    \advance\dimen@\pagetotal \advance\dimen@-\pageshrink
    \ifdim\dimen@>\pagegoal\@midfalse\p@gefalse\fi\fi
  \if@mid \ItemSep\box\z@\ItemSepbreak
  \else\insert\topins{\penalty100 
    \splittopskip\z@skip
    \splitmaxdepth\maxdimen \floatingpenalty\z@
    \ifp@ge \dimen@\dp\z@
    \vbox to\vsize{\unvbox\z@\kern-\dimen@}
    \else \box\z@\nobreak\ItemSep\fi}\fi\endgroup%
}


\def\gobbleone#1{}
\def\gobbletwo#1#2{}
\let\footnote=\gobbletwo 
\let\vfootnote=\gobbleone

\def\sing@footnote#1{\let\@sf\empty 
  \ifhmode\edef\@sf{\spacefactor\the\spacefactor}\/\fi
  \hbox{$^{\hbox{\eightpoint #1}}$}\@sf\sing@vfootnote{#1}%
}

\def\sing@vfootnote#1{\insert\footins\bgroup\eightpoint\b@ls{9pt}%
  \interlinepenalty\interfootnotelinepenalty
  \splittopskip\ht\strutbox 
  \splitmaxdepth\dp\strutbox \floatingpenalty\@MM
  \leftskip\z@skip \rightskip\z@skip \spaceskip\z@skip \xspaceskip\z@skip
  \noindent $^{\scriptstyle\hbox{#1}}$\hskip 4pt%
    \footstrut\futurelet\next\fo@t%
}

\def\footnoterule{\kern-3\p@ \hrule height \z@ \kern 3\p@}

\skip\footins=19.5pt plus 12pt minus 1pt
\count\footins=1000
\dimen\footins=\maxdimen


\def\landscape{%
  \global\TEMPDIMEN=\PageWidth
  \global\PageWidth=\PageHeight
  \global\PageHeight=\TEMPDIMEN
  \global\let\landscape=\relax
  \onecolumn
  \message{(landscape)}%
  \raggedbottom
}


\output{%
  \ifLeftCOL
    \global\setbox\LeftBOX=\vbox to \ZoneBSize{\box255\unvbox\ZoneBBOX}%
    \global\LeftCOLfalse
    \MakeRightCol
  \else
    \setbox\RightBOX=\vbox to \ZoneBSize{\box255\unvbox\ZoneBBOX}%
    \setbox\MidBOX=\hbox{\box\LeftBOX\hskip\ColumnGap\box\RightBOX}%
    \setbox\PageBOX=\vbox to \PageHeight{%
      \UnloadZoneA\box\MidBOX\UnloadZoneC}%
    \shipout\vbox{\PageHead\box\PageBOX\PageFoot}%
    \advancepageno
    \ifplate@page
      \shipout\vbox{%
        \sp@pagetrue
        \def\sp@type{plate}%
        \global\plate@pagefalse
        \PageHead\vbox to \PageHeight{\unvbox\plt@box\vfil}\PageFoot%
      }%
      \message{[plate]}%
      \advancepageno
    \fi
    \global\LeftCOLtrue
    \CleanStack
    \MakePage
  \fi
}


\Warn{\start@mess}


\catcode `\@=12 


 
\hoffset=-.5cm
\voffset=.5cm
\font\fivebmi=cmmib6
\font\sixbmi=cmmib6	\skewchar\sixbmi='177
\font\ninebmi=cmmib10 at 9pt 	\skewchar\ninebmi='177
\newfam\bmifam
\textfont\bmifam=\ninebmi
\scriptfont\bmifam=\sixbmi
\scriptscriptfont\bmifam=\fivebmi

\mathchardef\alpha="710B
\mathchardef\beta="710C
\mathchardef\gamma="710D
\mathchardef\delta="710E
\mathchardef\epsilon="710F
\mathchardef\zeta="7110
\mathchardef\eta="7111
\mathchardef\theta="7112
\mathchardef\iota="7113
\mathchardef\kappa="7114
\mathchardef\lambda="7115
\mathchardef\mu="7116
\mathchardef\nu="7117
\mathchardef\xi="7118
\mathchardef\pi="7119
\mathchardef\rho="711A
\mathchardef\sigma="711B
\mathchardef\tau="711C
\mathchardef\upsilon="711D
\mathchardef\phi="711E
\mathchardef\chi="711F
\mathchardef\psi="7120
\mathchardef\omega="7121
\mathchardef\varepsilon="7122
\mathchardef\vartheta="7123
\mathchardef\varpi="7124
\mathchardef\varrho="7125
\mathchardef\varsigma="7126
\mathchardef\varphi="7127

\def\chaphead{}
\newcount\eqnumber
\eqnumber=1

\def\today{\ifcase\month\or
 January\or February\or March\or April\or May\or June\or
 July\or August\or September\or October\or November\or December\fi
 \space\number\day, \number\year}

\def\eqnam#1{\xdef#1{(\chaphead\the\eqnumber}}

\def\newe{(\hbox{\chaphead\the\eqnumber})\global\advance\eqnumber by 1}
\def\firste{(\hbox{\chaphead\the\eqnumber a})\global\advance\eqnumber by 1}
\def\laste#1{\advance\eqnumber by -1%
	(\hbox{\chaphead\the\eqnumber #1})\advance\eqnumber by 1}

\def\refe#1{\advance\eqnumber by -#1 (\chaphead\the\eqnumber
     \advance\eqnumber by #1 }


\def\i{\relax\ifmmode{\rm i}\else\char16\fi}

\def\frac#1#2{{\textstyle{#1\over#2}}}

\def\d{{\rm d}}
\def\dddot#1{\ddot#1\kern-1.4pt\dot{\phantom{#1}}\kern-3pt}



\def\spose#1{\hbox to 0pt{#1\hss}}

\def\=#1{\overline{#1}}

\def\lta{\mathrel{\spose{\lower 3pt\hbox{$\mathchar"218$}}
     \raise 2.0pt\hbox{$\mathchar"13C$}}}
\def\gta{\mathrel{\spose{\lower 3pt\hbox{$\mathchar"218$}}
     \raise 2.0pt\hbox{$\mathchar"13E$}}}

\def\kpc{{\rm\,kpc}}

\def\msun{{\rm\,M_\odot}}
\def\lsun{{\rm\,L_\odot}}

\def\pc{{\rm\,pc}}

\def\annrev #1 #2 {ARA\&A, #1, #2}
\def\aa #1 #2 {A\&A, #1, #2}
\def\aasupp #1 #2 {A\&AS, #1, #2}
\def\aj #1 #2 {AJ, #1, #2}
\def\apj #1 #2 {ApJ, #1, #2}
\def\apjlett #1 #2 {ApJ, #1, #2}
\def\apjsupp #1 #2 {ApJS, #1, #2}
\def\ban #1 #2 {Bull.\ Astron.\ Inst.\ Netherlands, #1, #2}
\def\mn #1 #2 {MNRAS, #1, #2}
\def\nature #1 #2 {Nat, #1, #2}
\def\pasj #1 #2 {PASJ, #1, #2}
\def\pasp #1 #2 {PASP, #1, #2}

\input psfig

\overfullrule=0pt
\newif\ifpsfiles\psfilestrue

\def\frac#1#2{{#1\over#2}}
\def\label#1{}\def\cite#1{#1}\def\msun{\,{\cal M}_\odot}

\def\d{{\rm d}}
\def\cM{{\cal M}}\def\etal{{\it et al.}}
\Autonumber\begintopmatter
\title{The abundance of brown dwarfs}

\author{James Binney}
\affiliation{Theoretical Physics, University of Oxford, Oxford, OX1 3NP} 
\shortauthor{J.J.\ Binney}
\shorttitle{Brown dwarfs}

\abstract{The amount of mass contained in low-mass objects is investigated
anew. Instead of using a mass--luminosity relation to convert a luminosity
function to a mass function, I predict the mass-luminosity relation from
assumed mass functions and the luminosity functions of Jahreiss \& Wielen
(1997) and Gould, Bahcall \& Flynn (1997).  Comparison of the resulting
mass--luminosity relations with data for binary stars constrains the
permissible mass functions. If the mass function is assumed to be a power
law, the best fitting slope lies either side of the critical slope,
$\alpha=-2$, below which the mass in low-mass objects is divergent,
depending on the luminosity function adopted. If these power-law mass
functions are truncated at $0.001\msun$, the contribution to the local
density from stars lies between $0.016$ and $0.039\msun\pc^{-3}$, in
conformity with the mass density that has been inferred dynamically from
stars in the Hipparcos Catalogue. If the mass function is generalized from a
power law to a low-order
polynomial in $\log({\cal M})$, the mass in stars with ${\cal M}<0.1\msun$ is
either negligible or strongly divergent, depending on the order of the
polynomial adopted.
}

\keywords{Stars: IMF}

\maketitle

\section{Introduction}

The abundance in galaxies of low mass objects is of fundamental importance
because we know that near the Sun at least half of the Galaxy's mass density
is made up of stars fainter than $M_V=10$, and uncertainty in the expected
mass-to-light ratio of the Galactic disk is dominated by
uncertainty in the number of stars with $M_V\gta16$, which are extremely
hard to detect despite being intrinsically numerous.

The traditional way to determine the density of the lowest-mass stars is the
use of a wide-field proper-motion survey to pick up faint but nearby stars,
for which photometry and perhaps parallaxes can be obtained. More recently,
two alternative strategies have become available: (i) searches for gravitational
microlensing events, and (ii) narrow-field surveys
to identify extremely red stars. 

Microlensing surveys detect  stars through their gravitational
fields rather than their radiation, so microlensing surveys should provide a
powerful probe of the mass contained in low-mass stars. Unfortunately, there
are two large problems. First, to obtain even the projected density of
deflectors along lines of sight towards the survey stars, one must know how
the deflectors are distributed along the line of sight. Second, to determine
the mass function of the deflectors one requires a model of their
kinematics. In consequence of these difficulties, there is no concensus
regarding the nature of the deflectors that have caused the observed
microlensing events towards either the Galactic centre or the Magellanic
Clouds. 

The work of Gould and collaborators with HST (Gould, Bahcall \& Flynn, 1987;
Gould, Flynn \& Bahcall, 1988) pushes the strategy of counting extremely red
stars to its ultimate form, in which the limiting magnitude of the survey
becomes extremely faint and the field becomes very narrow.  Consequently,
most objects detected are of low luminosity and rather distant ($\sim2\kpc$)
(as are objects detected by microlensing surveys), and it is a non-trivial
task, that involves the adoption of a large-scale model of the Galaxy, to
infer the local luminosity function from the data. Despite these
difficulties, the luminosity function of the Galactic disk is now well
determined at $M_V\lta13$ and is usefully constrained down to $M_V\sim19$ --
see Fig.~1.

This paper is concerned with the problem of converting a known luminosity
function into a mass function.  The conventional procedure involves the
adoption of some mass--luminosity relation.  The mass--luminosity relation
for cool stars is complex and hard to determine either theoretically or
observationally, while we expect, a priori, that the mass function is
simple. Therefore, following Kroupa Tout \& Gilmore (1990) I {\it assume\/}
plausible mass functions and use measured luminosity functions to infer
mass--luminosity relations. Comparison of these inferred relations with the
data for binary stars clarifies the amount of mass that may reside in
sub-stellar objects. Section 2 explains the problem with the conventional
procedure, Section 3 applies an alternative approach to two possible
luminosity functions. Section 4 discusses the merits of the two luminosity
functions, and the implications of the mass functions for the local mass density.
Section 5 sums up.

\section{The problem}

Let  the number of stars in some volume of space that have masses in the
range $(\cM,\cM+\d\cM)$ be $\d N=\xi(\cM)\,\d\cM$. Then the number of stars that
have absolute magnitudes in the range $(M_V,M_V+\d M_V)$ is clearly
given by
 \eqnam\PhiXi$$
\Phi(M_V)={\d N\over\d M_V}=\xi(\cM)\left|{\d\cM\over\d M_V}\right|.
\eqno\newe$$
 The mass function $\xi$ is conventionally determined by estimating the
luminosity function $\Phi$ by counting stars, and then dividing equation
\refe1) through by $|{\d\cM\over\d M_V}|$. The problem with this procedure
is that  $|{\d\cM\over\d M_V}|$ goes to zero as one approaches the
hydrogen burning limit because there $M_V$ rises extremely rapidly with decreasing 
mass. Consequently, even the smallest error in the assumed value of 
$|{\d\cM\over\d M_V}|$ will give rise to a large error in the derived value
of $\xi(\cM)$.

This problem is exacerbated by the fact that $M_V(\cM)$ is not expected to
be a smooth function: the atmospheres of cool stars are extremely complex
with the result that their colours vary erratically with temperature and
therefore mass.  Moreover, low-luminosity stars take up to a Gyr to settle
onto the main sequence, so $M_V$ really needs to be treated as a function of
two variables. These considerations imply that any empirical determination
of $|{\d\cM\over\d M_V}|$, for example by fitting a polynomial to
empirically determined values of $M_V(\cM)$, will be unreliable.

Unfortunately, experts in the theoretical modelling of low luminosity stars
report that the prospects for determining $|{\d\cM\over\d M_V}|$
theoretically are no better. For example D'Antona \& Mazzitelli (1994) write
that {\it theoretical $T_{\rm eff}$'s, especially in the red, are
intrinsically ill-determined, and no sound observational interpretation
critically depending on the $T_{\rm eff}$'s can be presently performed.}
Clearly, any theoretical $M_V(\cM)$ relation does depends critically on
theoretical $T_{\rm eff}$'s, and the derivative of this relation will be
even more uncertain than the relation itself.

\section{An alternative approach}

The mass function, $\xi(\cM)$ is the outcome of a chaotic process, which
involves a wide range of densities, temperatures, velocities and magnetic
field strengths.  Consequently, it is much more likely to be featureless
than either the luminosity-mass relation, $M_V(\cM)$, or the luminosity
function, $\Phi(M_V)$, which is determined by $\xi(\cM)$ and $M_V(\cM)$.
Hence, it makes sense to assume a simple functional form for $\xi(\cM)$
and then to use the observed luminosity
function and equation \PhiXi) to predict $M_V(\cM)$ and to compare this
prediction to the relevant observational data.

\beginfigure1
\centerline{\psfig{file=PANEL1.PS,width=\hsize}}
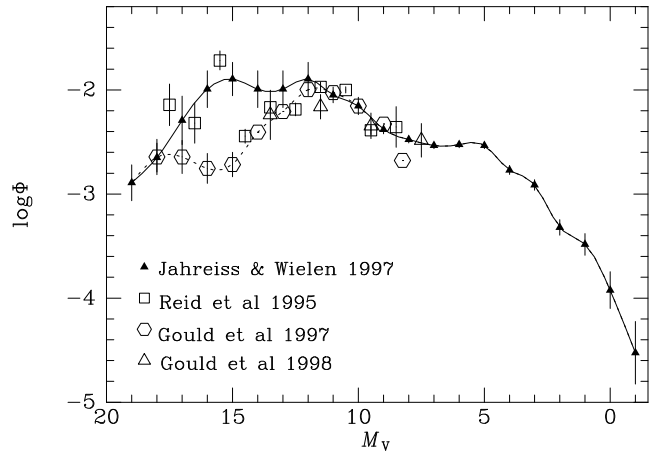
\caption{{\bf Figure 1.} The luminosity function of Jahreiss \& Wielen (1997)
(triangles) together with a cubic spline fit (full curve). Also
shown are the luminosity functions of Reid \etal\ (1995), Gould
\etal\ (1997), and Gould \etal\ (1998). 
The points  of Gould \etal\ (1997) have been shifted by $0.1\,$mag
to the left for clarity and joined by a cubic spline (dashed curve).}
\endfigure

\beginfigure2
\centerline{\psfig{file=PANEL2.PS,width=\hsize}}
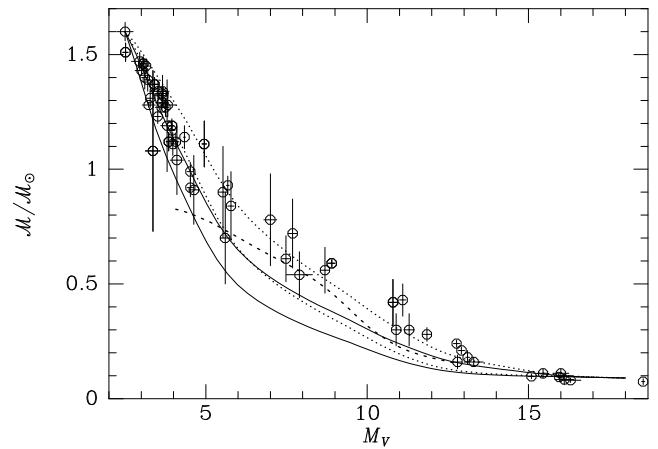
\caption{{\bf Figure 2.} The points show empirical 
determinations of the masses of binary
components from Popper (1980) and Henry \& McCarthy (1993). The dashed curve is
from Brewer \etal\ (1993) and indicates the predicted locus of subdwarfs. 
The full curves show the
mass-luminosity curve one obtains by assuming that $\xi\propto\cM^{-2.1}$.
The upper curve is for the Jahreiss \& Wielen luminosity function, while the
lower curve is for the Gould et al.\ (1997) luminosity function. The dotted
curves are obtained with the two luminosity functions and $\xi\propto\cM^{-2.1}$.}
\endfigure

From equation \PhiXi) we have (Kroupa et al.\ 1990)
$$
\int_{{\cal M}_1}^{{\cal M}_2}\d{\cal M}\,\xi=-\int_{M_1}^{M_2}\d M_V\,\Phi
\eqno\newe$$
With this equation  I use a generalization of the traditional
power-law mass function, namely
\eqnam\givesxi$$
\xi=\exp\Big(\sum_{n=0}^N\alpha_n\mu^n\Big),
\eqno\firste$$
 where
$$
\mu=\ln\big({\cal M}/\!\msun\big).
\eqno\laste b$$
For $N=1$ and $\alpha_1=-2.35$ this coincides with the Salpeter mass function.
For $N>1$ the mass function is not a simple power law and 
a characteristic mass is implied by each of the coefficients $\alpha_n$ for
$n>1$. 

Fig.~1 shows the two luminosity functions that I have used: the full curve
is a spline fit to  the luminosity function of Jahreiss \& Wielen (1997;
hereafter JW)
that employs Hipparcos parallaxes. An alternative luminosity function is
shown by the  dashed curve, which deviates from  the full curve to pass
through the points of  Gould, Bahcall \& Flynn (1997; hereafter GBF).

Fig.~2 shows four curves $M_V(\cM)$ that one obtains for the power-law case,
$N=1$, alongside empirical data points from Popper (1980) and Henry \&
McCarthy (1993). The two full curves are for $\alpha_1=-2.1$; the upper
curve is for the JW luminosity function, while the lower full
curve is for the GBF luminosity function. The two dotted
curves are for $\alpha_1=-1.8$.  The normalizations of $\Phi$ and the
integration constants in equation \refe2) have been chosen to force all
curves to coincide at $\cM=1.6\msun$ and $0.09\msun$.  The dashed curve,
which is from Brewer \etal\ (1993), is the result of fitting stellar models
to observed subdwarfs. Only one of these four curves can be said to be an
inadequate fit to the data, namely the lower full curve, which is for
$\alpha_1=-2.1$ and the GBF
luminosity function. In particular, for $M_V\lta4$ the upper full curve
($\alpha_1=-2.1$ and the JW luminosity function) fits the data better than
the upper dotted curve, that assumes the same luminosity function and
$\alpha=-1.8$.

Formally, none of the curves shown in Fig.~2 provides a satisfactory fit to
the data -- the lowest value of $\chi^2$ per degree of freedom is $10.2$,
which is attained for $\alpha_1=-2.1$ with the JW luminosity function.
Probably {\it no\/} plausibly smooth curve would give an acceptable fit to
the data, however, because at a given value of $M_V$ there are observational
points that lie far outside the error bars of other observational points,
probably because there is in reality no single mass-luminosity relation: age
and metallicity need to be taken into consideration. 

Although there is no compelling case to go beyond the simplest case $N=1$,
it is interesting to see what can be achieved by increasing $N$ to 2 and 3.
Fig.~3 shows the result of minimizing $\chi^2$ with respect to the
$\alpha_n$ in equation \givesxi a) in the cases $N=2$ (full curve) and $N=3$
(dotted curve). Both fits are for the JW luminosity function and give
significantly smaller values of $\chi^2$ per degre of freedom than any power
law: $\chi^2$ per degree of freedom is $4.1,\,3.1$ for $N=2,\,3$,
respectively, while the best-fiting power-law has $\alpha_1=-2.06$
and $\chi^2$ per degree of freedom $9.0$. (The best power-law with the
GBF luminosity function has $\alpha_1=-1.76$ and $\chi^2$ per degree of
freedom $\chi^2=13.3$.)

Fig.~4 shows the mass functions that underlie the fits of Fig.~3. These are
remarkably similar for $0.1\lta({\cal M}/\!\msun)\lta3$, which is the entire
range of masses over which the luminosity function contains information, and
are tangent to the best-fitting power law at ${\cal
M}=0.4\msun$. The negative curvature of these curves clearly implies that
the empirical mass-luminosity function is better fitted by a mass function
that at small masses turns down below the best-fitting power law.

\beginfigure3
\centerline{\psfig{file=PANEL3.PS,width=\hsize}}
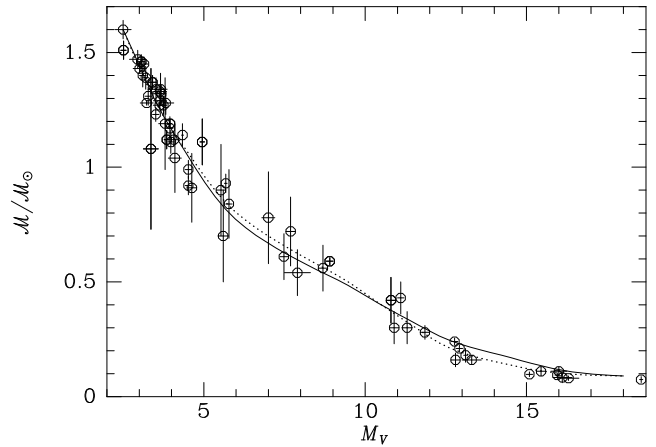
\caption{{\bf Figure 3.} The same as Fig.~2 except that the curves show the
result of minimizing $\chi^2$ with respect to the parameters $\alpha_n$ in
equation \givesxi a) for $N=2$ (full curve) and $N=3$ (dotted curve).
Both curves are based on the  JW luminosity function.}
\endfigure

\beginfigure4
\centerline{\psfig{file=PANEL4.PS,width=\hsize}}
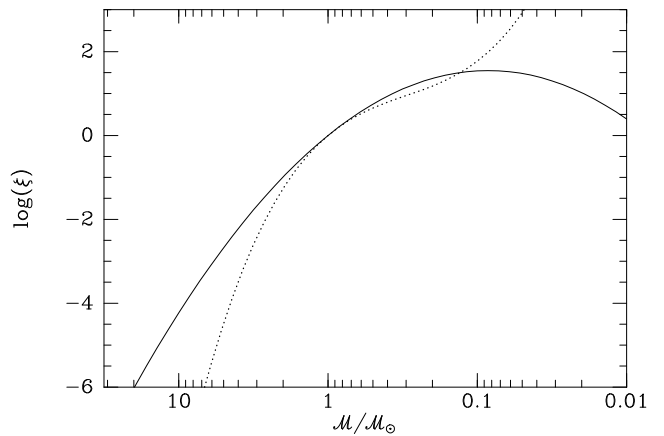
\caption{{\bf Figure 4.} The mass functions that correspond to the two
curves in Fig.~3.}
\endfigure

\section{DISCUSSION}

The luminosity function at $M_V\gta13$ is controversial.  Fig.~1 illustrates
this point by showing in addition to the JW points, values derived by Reid,
Hawley \& Gizis (1995) from a kinematically selected sample and by GBF and
Gould, Flynn \& Bahcall (1998) from photometrically selected samples of
stars observed with HST. The data of Gould, Flynn \& Bahcall, which are for
spheroid stars, only extend to $M_V=13.5$ and are compatible with the JW
points.  The points of Reid \etal\ and GBF go fainter and agree well down to
$M_V=14.5$.  GBF remark that had they been able to detect more secondaries
in binaries, their data would move up towards the Reid \etal\ points, which
have been corrected for unseen companions.  At $M_V=14.5$ the GBF and Reid
\etal\ data imply that $\Phi$ is a factor three smaller than the value of
JW. If the GBF luminosity function were to be preferred to that of JW,
Fig.~2 would rule out a simple power law with a slope $\alpha_1<-2$ that
predicts divergent integrated mass in low-mass objects. If the JW luminosity
function is correct, such values of $\alpha_1$ are not favoured over values
such as $\alpha_1=-1.8$ that place only finite mass in low-mass bodies.

The luminosity of a stellar population such as the Galactic disk is
dominated by stars with zero-age masses between $0.9\msun$ and $3\msun$.
Therefore, two stellar populations that have the same density of stars at
${\cal M}=2\msun$ and above will have very nearly the same luminosity,
regardless of their mass functions. Suppose two such populations have
power-law mass functions for ${\cal M}<2\msun$, one of slope $\alpha_1=-2.1$
and the other of slope $\alpha_1=-1.8$, and that in both cases the power law
is truncated at ${\cal M}=0.001\msun$. Then the mass, and hence the
mass-to-light ratio $\Upsilon$, of the population with $\alpha_1=-2.1$ will be
larger than that of the other population by a factor $2.9$. Hence the small
differences between acceptable curves in Fig.~2 are associated with
considerable differences in $\Upsilon$.

From Table 3.13 of Binney \& Merrifield (1998; hereafter BM) we take that a
$2\msun$ main-sequence star has $M_V=1.9$, and that at this mass $(\d
M_V/\d\ln{\cal M})=3.5$.  From Table 3.16 of BM we take that that there are
$4.8\times10^{-4}\pc^{-3}$ stars per cubic parsec with $1.4<M_V<2.4$. Hence,
with equation (1) a mass function that is normalized to produce the observed
density of $2\msun$ stars is
$$
\xi=
8.5\times10^{-4}\Big({{\cal M}\over2\!\msun}\Big)^{-\alpha}\msun^{-1}\pc^{-3}.
\eqno\newe$$
Integrating this down to $0.001\msun$ we find that the local mass density in
main-sequence stars is $0.039\msun\pc^{-3}$ if $\alpha=-2.1$ and
$0.013\msun\pc^{-3}$ if $\alpha=-1.8$. Since Table 3.16 of BM gives the
$V$-band luminosity density near the Sun as $0.053\lsun\pc^{-3}$, the
corresponding mass-to-light ratios are $\Upsilon_V=0.72$ and
$\Upsilon_V=0.25$. These numbers are usefully compared with the mass density
determined by Cr\'ez\'e et al (1998) from the Hipparcos catalogue:
$0.076\pm0.015\msun\pc^{-3}$. Of this latter value $\sim0.04\msun\pc^{-3}$
may be contributed by gas, and $\sim0.015\msun\pc^{-3}$ may be contained in
remnants. Hence the main-sequence density obtained with $\alpha=-2.1$ is on
the high side, and that obtained with $\alpha=-1.8$ is on the low side, but
neither number can be ruled out, especially given that the mean density of
local gas is very uncertain.

\section{Conclusions}

A priori we expect the mass function to be a smooth function, while we have
every reason to believe that the mass-luminosity relation is far from
smooth. Indeed, the whole concept of a mass-luminosity relation strictly
speaking fails for an inhomogeneous population such as that of the solar
neighbourhood because luminosity depends on age and metallicity in addition
to mass.  In these circumstances the optimum procedure is to assume a simple
functional form for the mass function and to determine the parameters in the
fitting function by calculating the mass-luminosity relation that is
obtained by combining it with the observed luminosity function. 

This has been done using binary data from Popper (1980) and Henry \&
McCarthy (1993) and two recent luminosity functions: that obtained by JW
from Hipparcos parallaxes, and that obtained by GBF from HST observations.
For both luminosity functions, acceptable fits to the data can be obtained
with a mass function that is a pure power law with exponent near the
critical value, $-2$, at which the mass implied at small masses diverges.
The JW luminosity function favours an exponent $\alpha_1=-2.06$ while the
GBF function favours $\alpha_1=-1.76$. If one continues a power-law mass
function that is normalized to yield the observed density of $\sim2\msun$
stars down to ${\cal M}=0.001\msun$, one obtains a local stellar mass
density $\rho_0=0.039\msun\pc^{-3}$ if $\alpha_1=-2.1$, and
$\rho_0=0.013\msun\pc^{-3}$ if $\alpha_1=-1.8$. The value inferred by
Cr\'ez\'e et al.\ (1998) from the dynamics of Hipparcos stars lies between
these figures depending on the rather uncertain value of the local density
of the ISM. Hence we cannot at the present time exclude the possibility that
most stellar mass is contained in stars less massive than the cutoff for
hydrogen burning.

Significantly better fits to the observed mass-luminosity relation can be
obtained by using a two- or three-parameter mass function rather than a
power-law. Such functions provide a better fit by virtue of their ability to
have a slope that decreases between $\msun$ and $0.1\msun$. If a
two-parameter function is employed, negligible mass is contained in stars
below the hydrogen-burning limit. By contrast,  a three-parameter function
predicts a huge amount of mass in low-mass stars. Consequently, nothing can
be securely inferred about the amount of mass in such stars from currently
available luminosity functions. The only real constraint is imposed by the
dynamics of nearby stars, and this constraint is not at all tight.

\section*{Acknowledgments} I am grateful to N.W.\ Evans and E.\ Kerins for
useful comments on a preliminary draft and to A.\ Gould for a valuable
referee's report.

\bye